\documentclass[pra,twocolumn]{revtex4-1}

\usepackage{graphicx}

\usepackage{bm}

\newcommand{\bvec}[1]{{\mathbf #1}}

\newcommand{\beq}{\begin{eqnarray}}
\newcommand{\eeq}{\end{eqnarray}}

\usepackage{color}

\begin{document}

\title{Phase and group velocities for correlation spreading in the Mott phase of the Bose-Hubbard model in dimensions greater than one}

\author{Ali Mokhtari-Jazi}
\author{Matthew R. C. Fitzpatrick}
\author{Malcolm P. Kennett}
\address{Department of Physics, Simon Fraser University, 8888 University Drive, Burnaby, British Columbia, V5A 1S6, Canada}

\date{\today}
\begin{abstract}
Lieb-Robinson and related bounds set an upper limit on the speed that information propagates in non-relativistic quantum systems.  Experimentally,
light-cone-like spreading has been observed for correlations in the Bose-Hubbard model (BHM) after a quantum quench. 
Using a two particle irreducible (2PI) strong-coupling approach to out-of-equilibrium dynamics in the BHM we calculate 
both the group and phase velocities for the spreading of single-particle correlations in one, two and three dimensions as a function of 
interaction strength.  Our results are in quantitative agreement with measurements of the speed of spreading of 
single-particle correlations in both the one- and two-dimensional BHM realized with ultra-cold atoms.  They also are consistent with with the claim
that the phase velocity rather than the group velocity was observed in recent experiments in two dimensions.  We demonstrate that there can be large differences between 
the phase and group velocities for the spreading of correlations and explore how  the anisotropy in the velocity varies
across the phase diagram of the BHM.  Our results establish the 2PI strong-coupling approach as a powerful
tool to study out-of-equilibrium dynamics in the BHM in dimensions greater than one.
\end{abstract}

\maketitle

\section{Introduction}

Ultra-cold atoms in optical lattices provide a versatile setting to investigate out-of-equilibrium dynamics in interacting quantum systems 
\cite{Greiner2002,Bloch2005,Lewenstein2007,Bloch2008,Hung2010,Chen2011,Kennett2013,Gross2017}. 
Atomic realizations of the Bose Hubbard Model (BHM)
\cite{Fisher1989}, a minimal model describing interacting bosons in an optical lattice \cite{Jaksch1998}, have also been proposed as quantum 
simulators in dimensions higher than one \cite{Choi2016,Takasu2020}. 
Understanding how information propagates in these systems provides insights that can
help the engineering of efficient quantum channels necessary for fast quantum computations \cite{Bose2007}. 
In this work we focus on the question of how the
speed of information propagation depends on dimensionality and model parameters in the BHM and whether theory can match 
experimental observations, particularly for two dimensions.

The existence of a bound on the group velocity of the spreading
of correlations in non-relativistic quantum spin systems with finite-range interactions
was demonstrated by Lieb and Robinson \cite{Lieb1972}.   For bosonic systems with infinite dimensional 
Hilbert space and unbounded Hamiltonians, analogous bounds can be found in some cases \cite{Nachtergaele2009},
but it is also possible to construct models with accelerating excitations \cite{Eisert2009}. Additionally, Lieb-Robinson 
bounds have been derived for spin-boson models relevant for trapped ions \cite{Juneman2013}.
In the specific case of the BHM, a bound has been derived for a specific class of initial
states 
\cite{Schuch2011}, but there are no rigorous results for 
the spreading of correlations in e.g. a Mott state.  Experimentally, in-situ imaging techniques such as quantum gas 
microscopes \cite{Bakr2010,Sherson2010} have enabled the demonstration of light-cone-like spreading \cite{Cheneau2012} of 
correlations for bosons in a one-dimensional optical lattice simulating the BHM.

There are multiple theoretical methods that enable the calculation of dynamical correlations in the BHM in one 
dimension, including exact diagonalization (ED) and time-dependent density-matrix renormalization group methods (t-DMRG)
\cite{Clark2004,Kollath2007,Lauchli2008,Bernier2011,Bernier2012,Barmettler2012,Cheneau2012,Trotzky2012,Cevolani2018,Despres2019}.
However, these tools are not effective for calculating the spreading of correlations in higher dimensions.
Theorists have responded to this challenge by using a variety of methods 
to study the spreading of correlations in the BHM in two dimensions, including
considering Gutzwiller mean-field theory with perturbative corrections \cite{Navez2010,Trefzger2011,Krutitsky2014,Quiesser2014}, time-dependent 
variational Monte Carlo \cite{Carleo2014}
and doublon-holon pair theories \cite{Yanay2016}.  We employ a two-particle irreducible (2PI) \cite{Cornwall1974,Berges2004} strong coupling approach to the
BHM developed by two of us that uses a closed time path method to treat out-of-equilibrium dynamics
(details can be found in Refs.~\cite{Kennett2011,Fitzpatrick2018a,Fitzpatrick2018b,Kennett2020} and Appendix \ref{sec:appA}) allowing accurate calculation
of the speed at which correlations spread in dimensions higher than one.  
This approach is exact in both the weak and strong interaction limits, and is applicable for small average particle number per site, $\bar{n}$.
In Ref.~\cite{Fitzpatrick2018a} two of us used this approach to obtain 2PI equations of motion for single-particle correlations.  
These equations of motion are not amenable to numerical solution due to the presence of multiple time integrals, but 
taking a low-energy limit yields an effective theory (ET) that gives predictions that match
exact results in one dimension \cite{Fitzpatrick2018b}.  

Our work is motivated by recent experiments reported by Takasu {\it et al.} \cite{Takasu2020}, which      
studied the spreading of 
single-particle correlations for bosonic atoms confined in an optical lattice in one, two and three dimensions after a quench 
in the optical lattice depth starting from a Mott insulating state. In terms of the BHM these quenches correspond to 
different values of the ratio $U/J_f$, where $U$ is the characteristic interaction energy scale and $J_f$ is the final value of the characteristic hopping energy scale $J$.
In one dimension they considered parameters well in the Mott phase [$U/J_f$ = 6.8, compared to the critical 
value $(U/J)^{1d}_c = 3.4$], and in two dimensions they considered parameters close to the 
transition to a superfluid [$U/J_f = 19.6$ compared to the critical value of $(U/J)^{2d}_c = 16$].
Defining the correlation wavefront as the first peak in the time evolution of the single-particle correlation function at each 
particle separation distance, Takasu {\it et al.} found the wavefront to propagate in one dimension with a velocity of $v_{\rm peak} = 5.5(7) Ja/\hbar$, where $a$ is the 
lattice spacing, in accord with previous experimental \cite{Cheneau2012} and theoretical results
\cite{Cheneau2012,Barmettler2012,Fitzpatrick2018b,Despres2019}.

Takasu {\it et al.} \cite{Takasu2020} reported the first measurements of propagation speeds in two dimensions, $v_{\rm peak} = 13.7(2.1) Ja/\hbar$ (obtained from the first peak in the 
single-particle correlations) and $v_{\rm trough} = 10.2(1.4) Ja/\hbar$ (obtained from the first trough 
after the first peak).  These values, especially $v_{\rm peak}$, are considerably larger than the value
 of  $v_{\rm max}^{2d} = 8.4 Ja/\hbar$ that Takasu {\it et al.} expected based on doublon-holon effective theories \cite{Cheneau2012,Barmettler2012}.
We note that Refs.~\cite{Cheneau2012,Barmettler2012} studied the one dimensional Bose Hubbard model and made use 
of the Jordan-Wigner transformation -- a one dimensional technique. 
The only derivation for $v_{\rm max}$, which Takasu {\it et al.} state can be interpreted as a Lieb-Robinson-like bound (i.e. corresponding to the group velocity) in dimensions larger than one that we are aware of is that of 
Krutitsky {\it et al.} \cite{Krutitsky2014}.  However, their
result agrees with the $v_{\rm max}^{2d}$ expression given by Takasu {\it et al.} only to zeroth order in $J/U$.  Moreover, the 
expression obtained by Krutitsky {\it et al.} is only valid deep in the Mott-insulating regime, so is not applicable to the $U/J$
values considered by Takasu {\it et al.} that are of interest here. 
Takasu {\it et al.} argue that they measured the phase rather than the group velocity; 
these were identified as being distinct for the BHM in one dimension in
Ref.~\cite{Despres2019}.

In this paper we solve the equations of motion for our 2PI ET, and calculate the group and phase velocities for the spreading of single particle correlations
in the BHM after a quench in one, two or three dimensions.

Our main results are: i) We obtain the group and phase velocities for correlation spreading throughout 
the Mott phase of the BHM in one, two and three dimensions; ii) We obtain quantitative agreement between the phase and group velocities of single-particle correlations 
in the one- and two-dimensional BHM calculated using our ET and velocities measured experimentally in Refs.~\cite{Cheneau2012,Takasu2020};
iii) We confirm that the phase rather than the group velocity was measured in two dimensions in Ref.~\cite{Takasu2020};
and iv) We track the evolution of anisotropy in the phase and group velocities in the BHM in both two and three dimensions. 

This paper is structured as follows: in Sec.~\ref{sec:model} we introduce the BHM and our methodology,
in Sec.~\ref{sec:results} we present our results for one, two and three dimensions, and in Sec.~\ref{sec:disc}
we discuss our results and conclude. 

\section{Model and methodology}
\label{sec:model}

We study the BHM on a $d$-dimensional cubic lattice, with $d = 1$, 2, and 3, for which the Hamiltonian is
\begin{eqnarray}
	\hat{H}_{\rm BHM} & = & -\sum_{\left<i,j\right>}J(t)\left(\hat{a}^\dagger_{\bvec{r}_i} \hat{a}^{\vphantom{\dagger}}_{\bvec{r}_j} + \hat{a}^\dagger_{\bvec{r}_j} 
	\hat{a}^{\vphantom{\dagger}}_{\bvec{r}_i}\right) - \mu \sum_i \hat{n}_{\bvec{r}_i}
	\nonumber \\ & & 
	+	\frac{U}{2}\sum_i \hat{n}_{\bvec{r}_i}\left(\hat{n}_{\bvec{r}_i} - 1\right),
\end{eqnarray}
where $\hat{a}_{\bvec{r}_i}^\dagger$ and $\hat{a}^{\vphantom{\dagger}}_{\bvec{r}_i}$ are bosonic creation and annihilation operators, respectively, 
and  $\hat{n}_{\bvec{r}_i}$ is the number operator, on site $i$ (located at $\bvec{r}_i$), $U$ is the interaction strength and $\mu$ the 
chemical potential. We restrict the hopping to be between nearest neighbour sites, and allow the magnitude $J(t)$ to be time 
dependent, as required for a quench protocol.  Our ET and the equations of
motion for the single-particle correlations were derived in Refs.~\cite{Fitzpatrick2018a} and \cite{Fitzpatrick2018b}.
We calculate the same quantity that was measured by Takasu {\it et al.}, the single-particle density matrix
\begin{eqnarray}
	\rho_1(\Delta\bvec{r} = \bvec{r}_i - \bvec{r}_j,t) & = & \left<\hat{a}^\dagger_{\bvec{r}_i}(t) \hat{a}^{\vphantom{\dagger}}_{\bvec{r}_j}(t)\right>,
\end{eqnarray}
 which contains all the information about single-particle observables,
and on a lattice can be written in the form
\begin{eqnarray}
 \rho_1(\Delta\bvec{r},t)  =  \frac{1}{N_s} \sum_{\bvec{k}} \cos\left(\bvec{k}\cdot\Delta\bvec{r}\right) n_{\bvec{k}}(t),
\end{eqnarray}
where $N_s$ is the number of sites, and $n_{\bvec{k}}(t)$ is the particle distribution over the quasi-momentum $\bvec{k}$ at time $t$,
and $\Delta\bvec{r}$ is the particle separation distance.  $n_{\bvec{k}}(t)$ is related to the density, $n(t)$, via
\begin{equation}
	n(t) = \frac{1}{N_s} \sum_{\bvec{k}} n_{\bvec{k}}(t).
\end{equation}
Note that usually one would expect the total particle number to be conserved, but due to truncations in our ET, there are 
small fluctuations in the particle number that do not appear to affect the determination of the velocity at which correlations spread \cite{Fitzpatrick2018b}. 
In addition to single-particle correlations, density-density correlations have also been considered in the literature \cite{Barmettler2012,Despres2019}.
Such correlations are not as easily accessible with our approach, but in the strong coupling limit of the BHM, higher order correlations
contain the same information as single-particle correlations \cite{Barmettler2012,Fitzpatrick2018b}.

The protocol we follow is to start with $J/U = 0$ for a $\bar{n} = 1$ Mott phase and then ramp $J$ to a final value $J_f$ over a timescale
$\tau_Q$, with the timescale $t_c$ marking the midpoint of the quench  \cite{Fitzpatrick2018b}.  We solve the ET equations of motion to obtain $\rho_1(\Delta\bvec{r},t)$,
 from which we extract the group and phase velocities for the spreading of single-particle correlations.
 More details on the procedure we used can be found in Appendix \ref{sec:appB}.
We now discuss our results for one, two and three dimensions in turn. 

\begin{figure}[ht]
    \includegraphics[width=8cm]{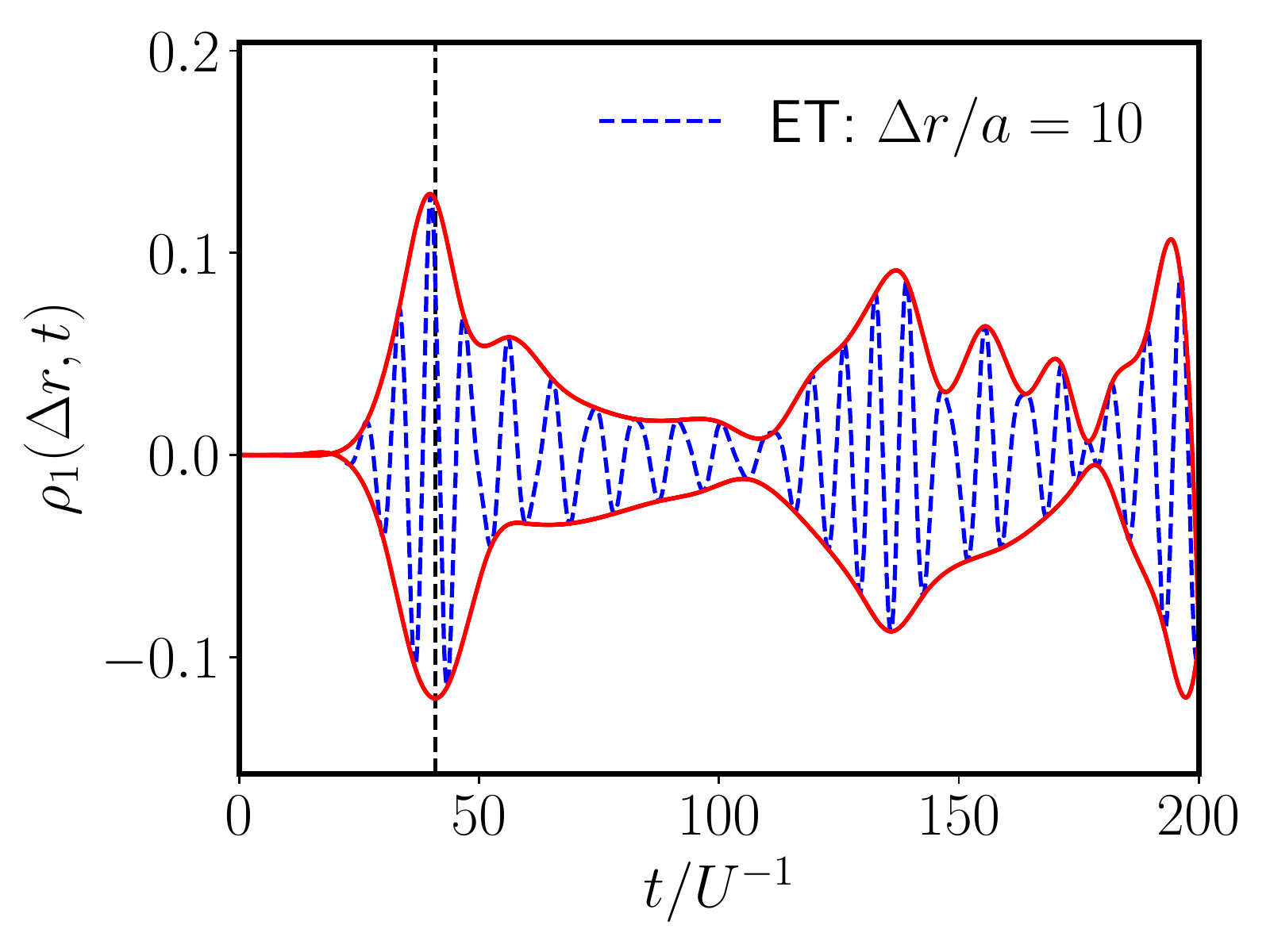}
\caption{Dynamics of $\rho_1(\Delta r,t)$ calculated from the ET in one dimension for $\Delta r/a = 10$. The envelope of the wavepacket is shown in red
and the centre of the wave packet is marked by the dashed black vertical line.
We use parameters $\beta U = 1000$, $U/J_f = 18.2$, $\mu/U = 0.4116$, $t_c /U^{-1} = 5$, $t_Q/ U^{-1} = 0.1$ and $N_s = 50$.}
\label{fig:1d_rho1_a}
\end{figure}

\begin{figure}[th]
    \includegraphics[width=8cm]{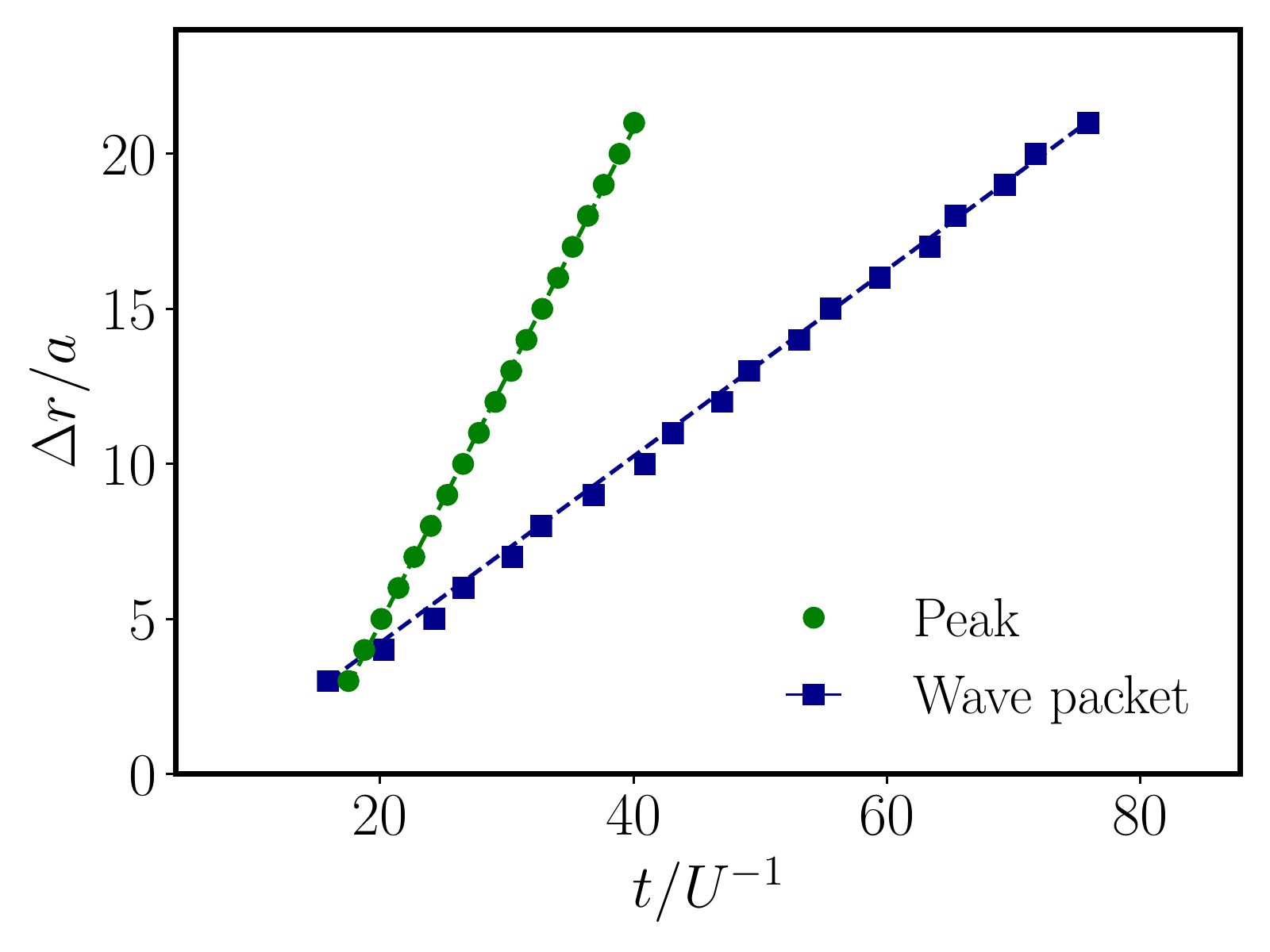}
        \caption{Scatterplots of the time $t/U^{-1}$ for the maximum peak (green) and the wave packet (blue) to travel a distance $\Delta \bvec{r}/a$.
Parameters are the same as in Fig.~\ref{fig:1d_rho1_a}.}
\label{fig:1d_rho1_b}
\end{figure}

\section{Results}
\label{sec:results}
\subsection{One dimension}
We consider one-dimensional chains with 50 sites and periodic boundary conditions (PBC).  In previous work \cite{Fitzpatrick2018b} we showed that the spreading of 
correlations calculated with our ET matches well with ED results in small systems and exact results for 
larger systems in one dimension.
For a given $U/J_f$, we calculate $\rho_1(\Delta\bvec{r},t)$  and for each value of $\Delta \mathbf{r}$ we obtain the timewise positions of the wave packet, 
and the largest peak [i.e. the point in time where $\rho_1\left(\Delta\mathbf{r}, t\right)$ takes its maximum value]. 
We track the propagation of the maximum peak of the $\rho_1\left(\Delta\mathbf{r}, t\right)$ time series to extract the phase velocity.  To obtain 
the group velocity we first perform both linear and cubic interpolations to determine the upper 
and lower envelopes of $\rho_1\left(\Delta \mathbf{r}, t\right)$ and then 
average the centres of the pairs of upper and lower envelopes to identify  the position of the wave packet.
Full details of our procedure are given in Appendix \ref{sec:appB}.
An example of the envelope for $\rho_1\left(\Delta\mathbf{r}, t\right)$
is given in Fig.~\ref{fig:1d_rho1_a} for $\Delta r/a=10$ and $U/J_f=18.2$, with the timewise position of the wave packet marked by a vertical dashed black line.
By tracking the propagation of the wave packet, we can extract the group velocity 
for the spreading of single-particle correlations \cite{Fitzpatrick2018b}. 
Figure~\ref{fig:1d_rho1_b} plots the times $t/U^{-1}$ for the maximum peak and the wave packet to travel a particle separation distance $\Delta r/a$ for the
same parameters used in Fig.~\ref{fig:1d_rho1_a}. By performing linear fits to the data in Fig.~\ref{fig:1d_rho1_b}, we extract estimates for the phase and group velocities.
The wavepackets show less damping than those seen experimentally \cite{Cheneau2012,Takasu2020}.
There are several possible sources for this discrepancy: on the theoretical side, there are
higher order terms in the 2PI expansion that lead to an imaginary part of the self-energy which should lead to
damping \cite{Stefanucci2013}.  We exclude these terms as they should be small in comparison to the terms that we keep
and in doing so the problem becomes more numerically tractable.  Experimentally, sources of damping
that are not included in our calculation, such as atom number fluctuations may also be important. 

\begin{figure}[ht]
 \centering
 \includegraphics[width=0.41\textwidth]{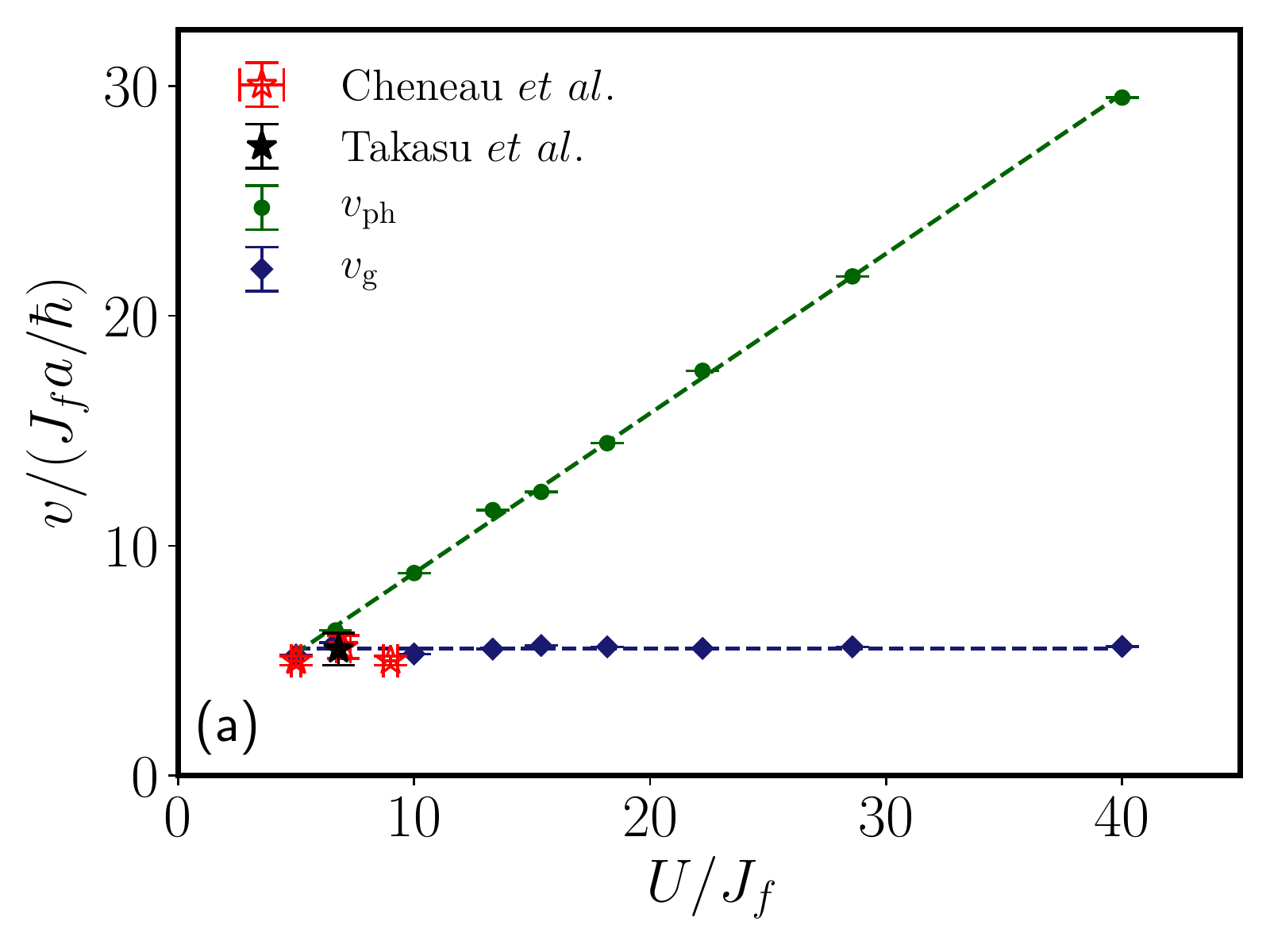}
 \includegraphics[width=0.41\textwidth]{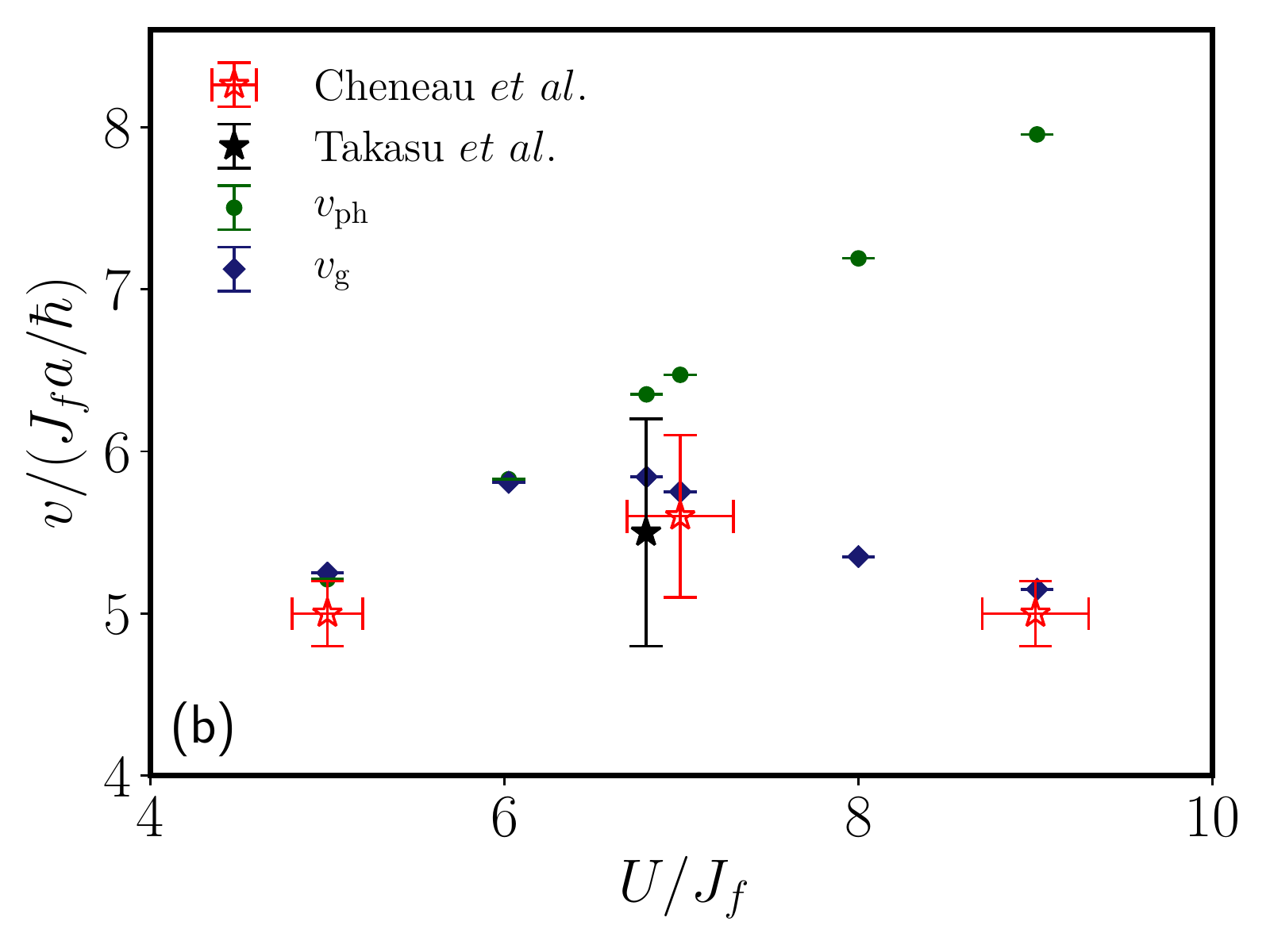}
	\caption{(a) Phase velocity ($v_{\rm ph}$) and group velocity ($v_{\rm g}$) for single particle correlations as a function of $U/J_f$ for a 
	50 site chain.  Experimental results from Cheneau {\it et al.} \cite{Cheneau2012} and Takasu {\it et al.} \cite{Takasu2020} are also shown. The dashed lines are guides to the eye. 
	(b) Comparison between experiment and theory in the range of $U/J_f$ where experimental data is available.}
  \label{fig:1d_group_phase}
\end{figure}

We repeat the process illustrated in Figs.~\ref{fig:1d_rho1_a} and \ref{fig:1d_rho1_b} throughout the Mott phase to determine the group and phase velocity 
at each value of $U/J_f$.  These results and a comparison to the velocities determined experimentally in Refs.~\cite{Takasu2020} and \cite{Cheneau2012} are presented in Fig.~\ref{fig:1d_group_phase}.
Note that the velocities obtained in Ref.~\cite{Cheneau2012} are actually for density-density correlations rather than single-particle correlations, but at strong coupling,
these two correlations should spread with similar velocities \cite{Barmettler2012,Fitzpatrick2018b}.
Deep in the Mott insulating phase, the phase velocity is much larger than the group velocity but the two velocities converge in the vicinity of the critical point.  Our results
are consistent with those recently obtained theoretically using matrix product states by Despres {\it et al.}~\cite{Despres2019}.  Our results for the group velocity 
are consistent with measurements of the group velocity by Cheneau {\it et al.}~\cite{Cheneau2012} at several different values of $U/J_f$.  Takasu {\it et al.} \cite{Takasu2020} identified
their results with the group velocity, which agree with the value we compute for the group velocity.  However, they used the position of the peak to determine the 
velocity, which would suggest they may have measured the phase velocity -- they note that the two values are quite close for $U/J_f = 6.8$.  The value we obtain for
the phase velocity is just outside the error bars of Takasu {\it et al.}'s measurement, but given that we consider a uniform system, whereas the experiment is in a 
trap, it would appear that Takasu {\it et al.}'s results are not inconsistent with them having measured the phase velocity in one dimension. 
Having established that our ET reproduces existing experimental results for the group velocity and theoretical results obtained using essentially exact methods in one dimension, we now turn to 
two dimensions, where the phase velocity has not been previously considered theoretically.

\begin{figure}[th]
    \includegraphics[width=7cm]{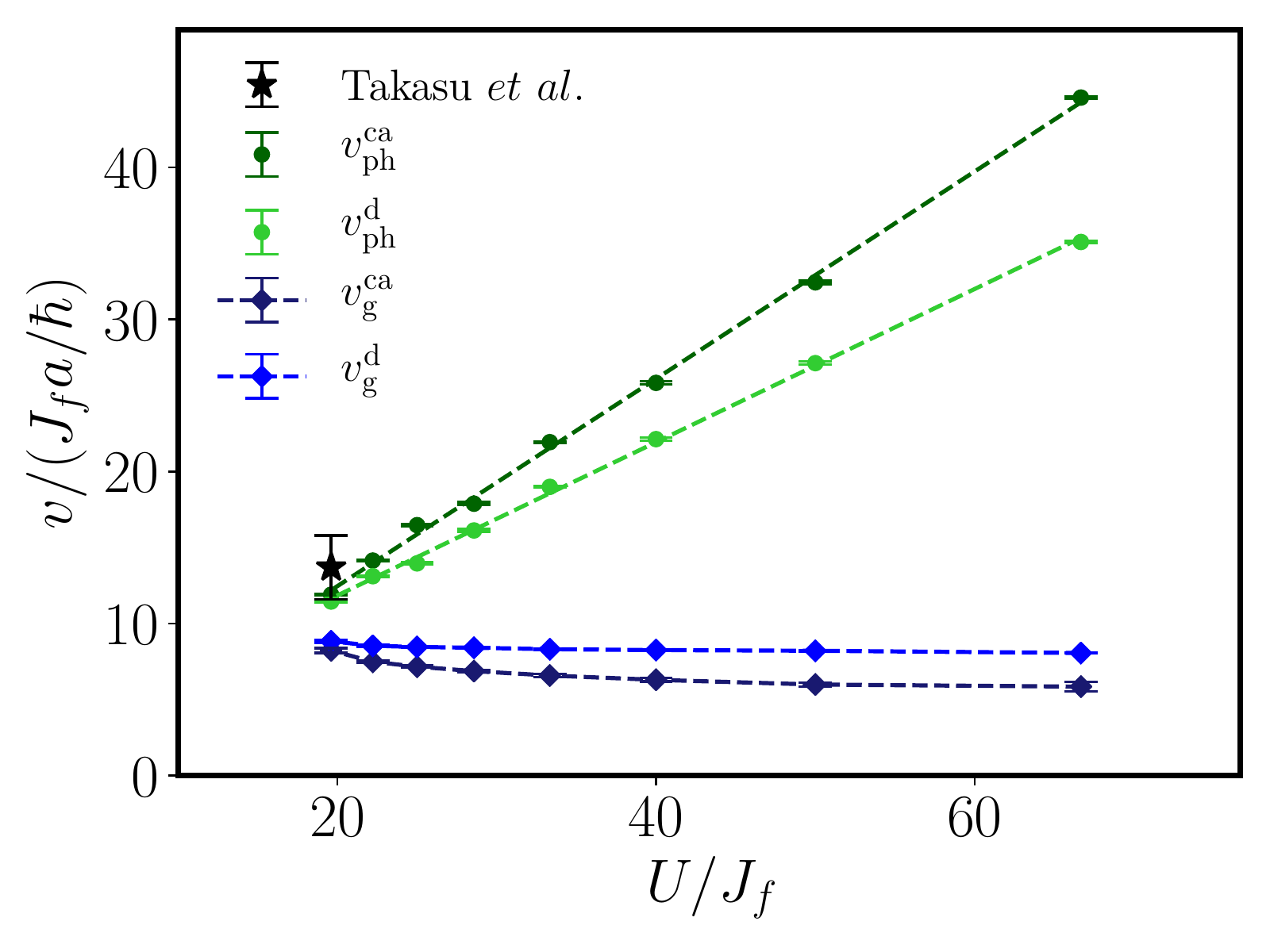}
        \caption{Phase velocity ($v_{\rm ph}$) and group velocity ($v_{\rm g}$) along both the crystal axes (superscript ca) and along the diagonal (superscript d) as
        a function of $U/J_f$ for the BHM in two dimensions.
         The peak velocity in the single-particle correlations determined by Takasu {\it et al}. \cite{Takasu2020} is also shown.}
\label{fig:2d_group_phase_velocity}
\end{figure}

\subsection{Two dimensions}
We consider a 50 $\times$ 50 lattice with PBC and follow the same procedure as for one dimension to calculate $\rho_1(\Delta {\bvec{r}},t)$.
As noted by previous authors, the spreading of correlations in two dimensions is anisotropic in both the Mott \cite{Krutitsky2014,Fitzpatrick2018b}
and superfluid \cite{Carleo2014} regimes.  We calculate the phase and group velocities as a function of $U/J_f$ along both the 
crystal axes and the diagonals using the same protocol as for one dimension and present the results in Fig.~\ref{fig:2d_group_phase_velocity}.  
We consider parameters $\beta U = 1000$, $\mu/U = 0.4116$, $t_c /U^{-1} = 5$, and $t_Q/ U^{-1} = 0.1$

Takasu {\it et al.} \cite{Takasu2020} identified propagation velocities for single-particle correlations by fitting to the peak and the following trough in $\rho_1(\Delta {\bvec{r}},t)$ at each
$\Delta r$.  These values were $v_{\rm peak} = 13.7(2.1) J_f a/{\hbar}$ and $v_{\rm trough} = 10.2(1.4) J_f a/{\hbar}$ for $U/J_f = 19.6$. In Fig.~\ref{fig:2d_group_phase_velocity} we show
the group and phase velocities, evaluated along both the diagonals and the crystal axes in two dimensions. We used the peaks in $\rho_1(\Delta {\bvec{r}},t)$ to calculate the phase velocity and
accordingly plot Takasu {\it et al.}'s $v_{\rm peak}$ value in Fig.~\ref{fig:2d_group_phase_velocity}.  For $U/J_f = 19.6$ we find the group velocity and phase velocity (determined using the peak in  $\rho_1(\Delta {\bvec{r}},t)$)
along the diagonals to be $v^{\rm d}_{\rm g} \simeq 8.8 J_f a/{\hbar}$ and $v^{\rm d}_{\rm ph} \simeq 11.5 J_f a/{\hbar}$
and along the crystal axes  $v^{\rm ca}_{\rm g} \simeq 8.2 J_f a/{\hbar}$ and $v^{\rm ca}_{\rm ph} \simeq 11.9 J_f a/{\hbar}$ respectively. 
We also determined the phase velocity using the first trough after the peak in
$\rho_1(\Delta {\bvec{r}},t)$ and obtained $v^{\rm ca}_{\rm ph} = 11.2 J_f a/{\hbar}$ and $v^{\rm d}_{\rm ph} = 11.0 J_f a/{\hbar}$, also consistent with experiment.  This result indicates a strength of our method relative to the truncated Wigner approximation (TWA) used in Ref.~\cite{Takasu2020}, which failed to capture the locations of the correlation troughs.
Our results are consistent with Takasu {\it et al.}'s statement that they measure the phase velocity rather than the group velocity for the 
spreading of correlations. 

\begin{figure}[th]
        \includegraphics[width=9cm]{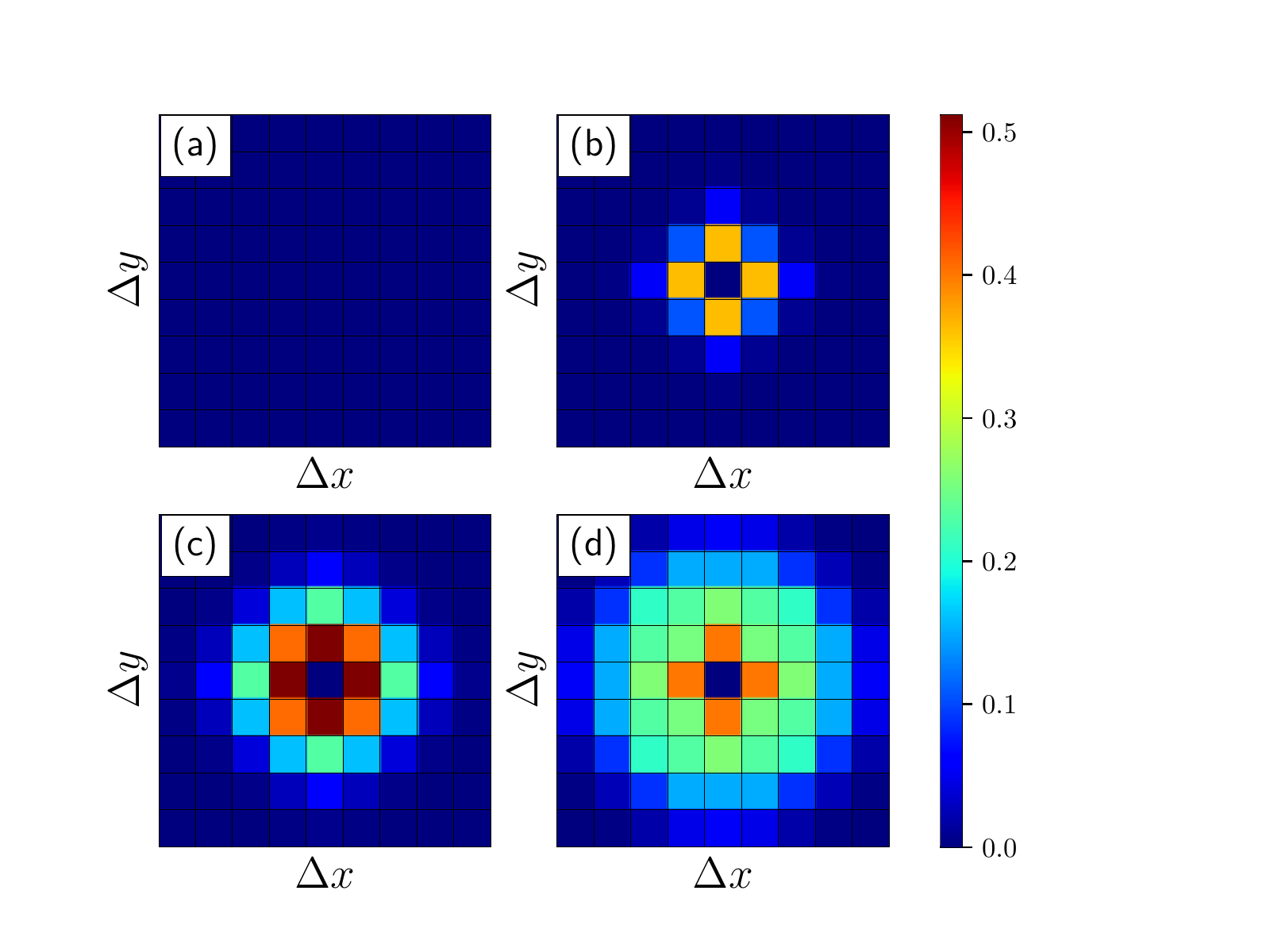}
        \caption{Snapshots of $\rho_1(\Delta {\bvec{r}},t)$ at four times after the quench:
        (a) 0, (b) 0.12 $\hbar/J$, (c) 0.23 $\hbar/J$, and (d) 0.35 $\hbar/J$. The parameters used are $\beta U = 1000$,
        $U/J_f = 19.6$, $\mu/U = 0.4116$ and $N_s = 50$.}
\label{fig:2d_spreading}
\end{figure}

We find both the group and phase velocities to be anisotropic, but with opposite sense -- the velocity along the diagonals is larger than along the crystal axis for the group velocity and the converse for the phase velocity.
The degree of anisotropy of both the phase and group velocities in two dimensions is maximum at large values of $U/J_f$ and the velocities are close to 
isotropic  as $U/J_f$ approaches the critical value of  $(U/J)^{2d}_c = 16$.  Theoretical calculations for the group velocity in the superfluid regime \cite{Carleo2014} indicate that the superfluid also
displays anisotropic spreading of correlations, with the opposite sense to that in the Mott regime.  
In Fig.~\ref{fig:2d_spreading} we show the spreading of $\rho_1(\Delta {\bvec{r}},t)$ at four times after the quench, for 
Euclidean distances $\Delta \leq 4$, as measured by Takasu {\it et al.} (we display correlations for the same times as those shown in Ref.~\cite{Takasu2020}).  
The magnitude of  $\rho_1(\Delta {\bvec{r}},t)$ we find from our ET is about twice the amplitude at the peak observed in experiment or ED
 calculations in small systems, but the phase of $\rho_1(\Delta {\bvec{r}},t)$ appears to be considerably more accurate.  
 At larger values of $U/J_f$ the ET accurately reproduces ED results as illustrated in Appendix \ref{sec:appB}.

\begin{figure}[th]
    \includegraphics[width=7cm]{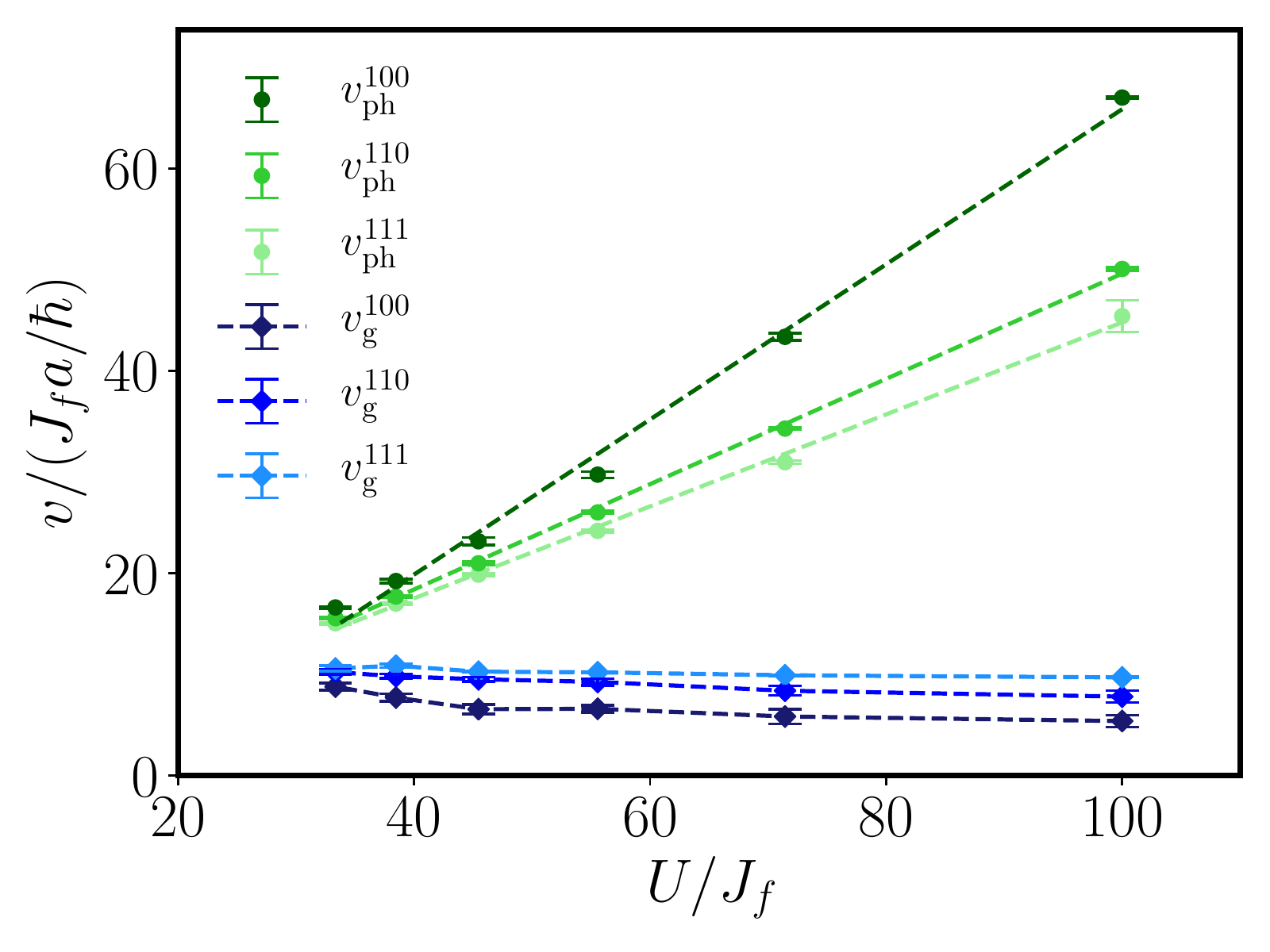}
        \caption{Phase velocity ($v_{\rm ph}$) and group velocity ($v_{\rm g}$)  along the (1,0,0), (1,1,0) and (1,1,1) directions as a function of $U/J_f$
          for the BHM in three dimensions.}
\label{fig:3d_group_phase_velocity}
\end{figure}

\subsection{Three dimensions}
We followed a similar procedure to one and two dimensions to determine the group and 
phase velocities for the spreading of correlations in three dimensions for a 28$\times$28$\times$28 lattice with PBC, and the results are illustrated in 
Fig.~\ref{fig:3d_group_phase_velocity}.  The group velocity
is relatively insensitive to $U/J_f$, as found in Ref.~\cite{Fitzpatrick2018b}, whereas the phase velocity increases approximately linearly with $U/J_f$.
Similarly to two dimensions, there is also anisotropy in both the group and phase velocities which decreases as the critical point is approached,
and it has the opposite sense for phase and group velocities.  The group velocity is maximal along the (1,1,1) direction and the phase velocity is minimal along 
the same direction at a given $U/J_f$.

\section{Discussion and conclusions}
\label{sec:disc}
We have applied our 2PI strong coupling approach to the BHM to calculate the spreading of single-particle correlations and found
excellent agreement with experiments in one \cite{Cheneau2012,Takasu2020} and two \cite{Takasu2020} dimensions.  This establishes our 2PI strong 
coupling approach as a powerful tool to study out-of-equilibrium dynamics in the BHM in dimensions greater than one.  Given that the method gives 
more accurate results for equilibrium properties, such as phase boundaries, with increasing dimension \cite{Fitzpatrick2018a}, we expect the same to 
be true for out-of-equilibrium dynamics.  Hence, as it reproduces exact results in one dimension, the 2PI method is complementary to numerical 
methods that give essentially exact results for out-of-equilibrium dynamics only in one dimension.  In addition, the 2PI method can be extended to disordered 
systems \cite{Fitzpatrick2019,Kennett2020} and multi-component boson systems.  We have also demonstrated anisotropy
in the spreading of correlations on a lattice in both two and three dimensions.  This anisotropy persists throughout the entire Mott phase, apparently 
vanishing only around the critical point.  Our results for the phase velocity and group velocity as a function of $U/J_f$ demonstrate that
while they are relatively similar in the vicinity of the transition to the superfluid, deeper in the Mott phase there can be very significant 
differences, with the phase velocity being much larger than the group velocity.  Differentiating between these two velocities is important
for understanding the rate of information spreading in the BHM. 
At the present time there have been only a few measurements of the velocities at which correlations spread in the BHM,
we hope that our results give an incentive to further experimental measurements of correlation spreading in the BHM.

\section{Acknowledgements}
A. M-J. and M. K. acknowledge support from NSERC, and M. F. was supported by a Mitacs Accelerate Postdoctoral Fellowship.  The authors
thank Y. Takasu for helpful correspondence and J. McGuirk and R. Wortis for thoughtful comments on the manuscript.

\appendix

\section{Real-time 2PI approach to the Bose-Hubbard Model}
\label{sec:appA}
In this appendix we provide a brief summary of our real-time two-particle irreducible (2PI) approach to the Bose-Hubbard model
developed in  Refs.~\cite{Fitzpatrick2018a} and \cite{Fitzpatrick2018b}).
We then give a brief review of the equations of
motion we solve.

\begin{figure}[h]
    \includegraphics[width=8cm]{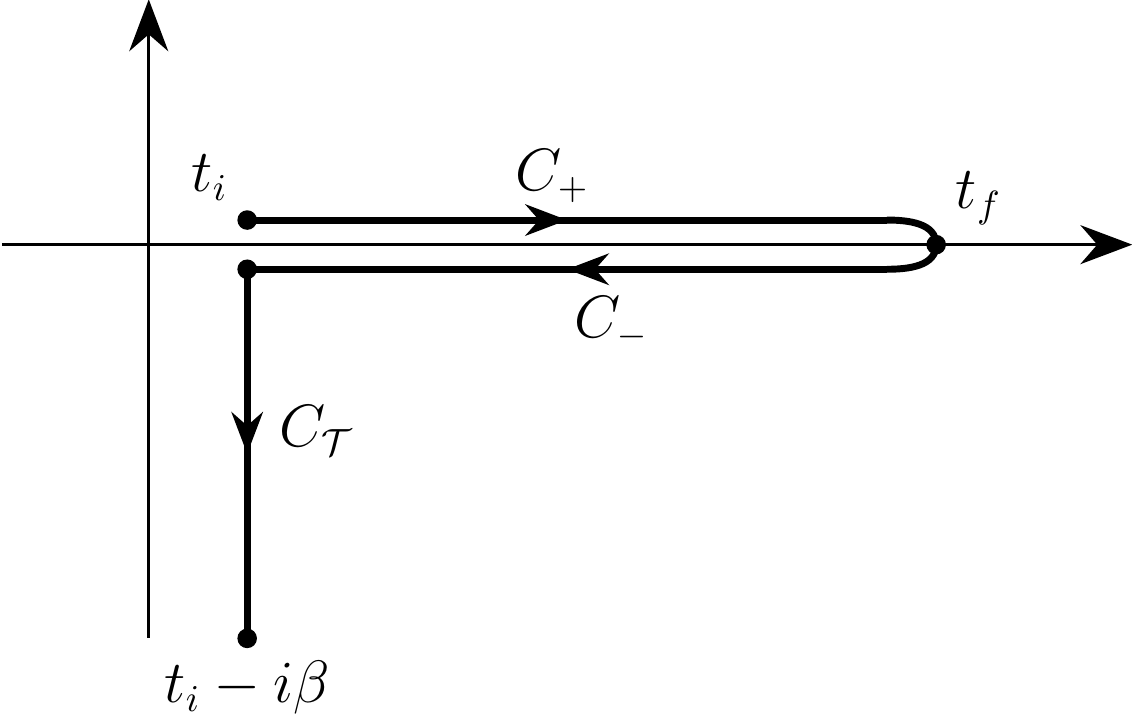}
\caption{KP contour used in calculations.  The system is taken to be prepared in an initially thermal state at time $t_i$ and then evolved to a final
        time $t_f$.}
\label{fig:contour}
\end{figure}

We use a contour-time formalism, making use of the Konstantinov-Perel (KP) \cite{Konstantinov1961} contour, illustrated in Fig.~\ref{fig:contour}.
Fields are labelled according to their contour, which can be $+$, $-$ or ${\mathcal T}$, and we use the notation $\hat{a}^a_{\bvec{r}_i\alpha}$ for bosonic fields on site $i$ 
(located at position $\bvec{r}_i$)
and contour $\alpha$ where $\hat{a}^1_{\bvec{r}_i\alpha} = \hat{a}^{\vphantom{1}}_{\bvec{r}_i\alpha}$ and $\hat{a}^2_{\bvec{r}_i\alpha} = \hat{a}^\dagger_{\bvec{r}_i\alpha}$.  We cast the generating functional
${\mathcal Z}$ for the Bose-Hubbard model in path integral form (omitting source terms)
\begin{equation} 
	{\mathcal Z} = \int [{\mathcal D}a] e^{iS_{\rm BHM}[a]},
\end{equation}
where $S_{\rm BHM}[a]$ is the action for the Bose-Hubbard model. 

We perform two Hubbard-Stratonovich transformations on the BHM action \cite{Sengupta2005,Kennett2011,Fitzpatrick2018a,Fitzpatrick2018b} 
and then perform a cumulant expansion to quartic order, which gives the following action \cite{Fitzpatrick2018a}:
\begin{widetext}
\begin{equation}
 S_{\rm BHM}[z] = \frac{1}{2!} \left( 2J_{x_1 x_2} + \left[{\mathcal G}^{-1}\right]_{x_1 x_2} + \tilde{u}_{x_1 x_2}\right) z_{x_1} z_{x_2}
+ \frac{1}{4!} u_{x_1 \ldots x_{4}} z_{x_1} z_{x_2} z_{x_3} z_{x_{4}},
	\label{eq:effective_theory}
\end{equation}
\end{widetext}
where the $z$ fields can be shown to have the same correlations as the original $a$ fields \cite{Sengupta2005,Fitzpatrick2018a}. 
We make use of the Einstein summation convention and introduce highly condensed notation so that for a quantity $Q$
\begin{equation}
	Q_{x_1 x_2 \ldots x_n} = Q^{a_1 a_2 \ldots a_n}_{\bvec{r}_{i_1} \bvec{r}_{i_2} \ldots \bvec{r}_{i_n}, \alpha_1 \alpha_2 \ldots \alpha_n}(s_1, s_2,
        \ldots, s_n),
\end{equation}
where $\alpha_i$ is a contour label, $\bvec{r}_{i_n}$ is a lattice site position, and $s_i$ is a non-negative real parameter that indicates a position along the 
contour segment $C_{\alpha_i}$.
The $u$ vertex is non-local in time and generates ``physical'' and ``anomalous'' diagrams (i.e. those with internal lines of ${\mathcal G}^{-1}$).
The $\tilde{u}$ vertex cancels anomalous terms generated by $u$ vertices, and
${\mathcal G}$ and ${\mathcal G^{2}}$ are connected 1 and 2 particle Green's functions
in the atomic limit ($J=0$).  These vertices are illustrated in Fig.~\ref{fig:vertices}.
Full details of the expressions for each of these vertices are presented in Ref.~\cite{Fitzpatrick2018a}.
With the effective theory defined by Eq.~(\ref{eq:effective_theory}) we can construct the 2PI effective action from
whence we obtain equations of motion.

\begin{widetext}

\begin{figure}[h]
        \begin{center}
\includegraphics[height=4cm]{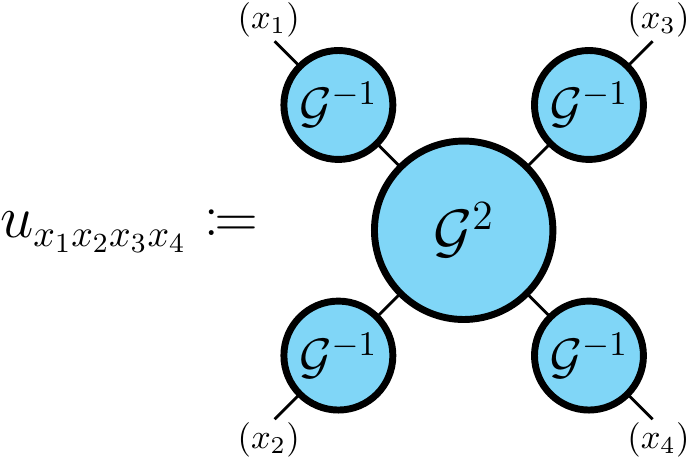} \quad \quad \quad
\includegraphics[height=4cm]{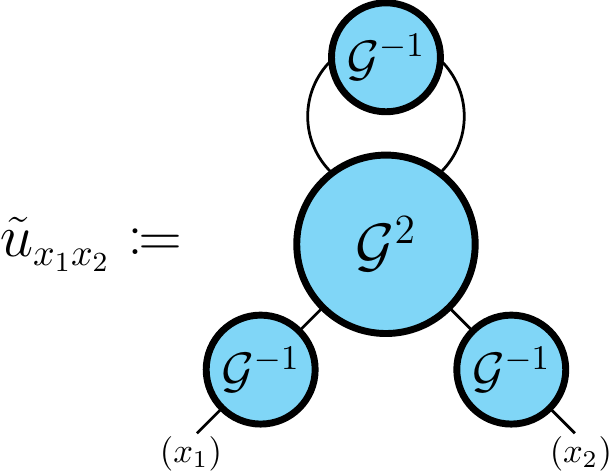}
\end{center}
\caption{The four-point vertex $u$ and the two-point vertex $\tilde{u}$ in the effective theory.}
\label{fig:vertices}
\end{figure}

\begin{figure}[h]
        \begin{center}
           \includegraphics[width=10cm]{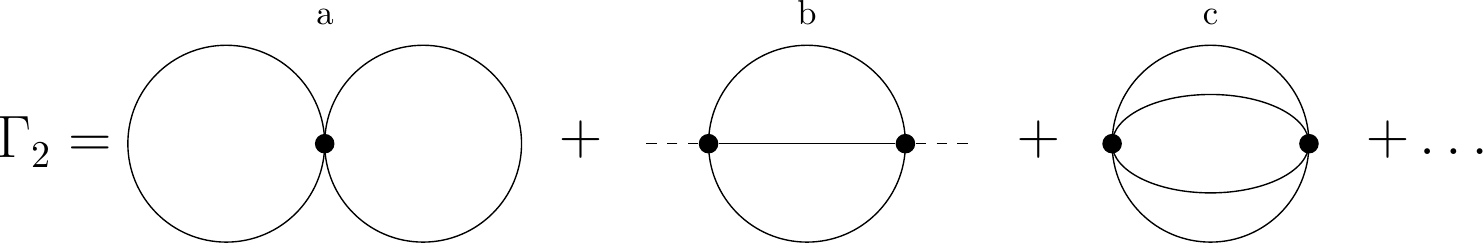}
        \end{center}
\caption{Diagrammatic expansion of $\Gamma_2$. The interaction vertex $u$ is represented by a bullet: $\bullet$, the propagator
$G$ is represented by a solid line and the superfluid order parameter $\phi$ is represented by a dashed line.}
\label{fig:gamma2}
\end{figure}
\end{widetext}

\subsection{2PI equations of motion}
With the effective theory defined by Eq.~(\ref{eq:effective_theory}) we can construct the 2PI effective action, and hence obtain the 
2PI equations of motion \cite{Cornwall1974}. The equations of motion for the superfluid order parameter $\phi$ and the propagator $G$ take the form
\begin{equation}
        \frac{\delta S[\phi]}{\delta \phi_{x}} + \frac{1}{2} \left[i \frac{\delta \left[D^{-1}\right]_{x_1 x_2}}{\delta \phi_{x}} G_{x_2 x_1} \right] 
        + \frac{\delta \Gamma_2[\phi,G]}{\delta \phi_{x}}  = 0,
\label{eq:eqm1}
\end{equation}
and
\begin{equation}
        i \left[G^{-1}\right]_{x_1 x_2}  = i \left[D^{-1}\right]_{x_1 x_2} - i \Sigma_{x_1 x_2}, 
\label{eq:eqm2}
\end{equation}
where Eq.~(\ref{eq:eqm2}) is the Dyson's equation, $i \left[D^{-1}\right]$ is the inverse ``bare'' propagator and $\Sigma$ is the 2PI self-energy:
\begin{equation}
        i \left[D^{-1}\right]_{x_1 x_2} 
        \equiv \frac{\delta^2 S[\phi]}{\delta \phi_{x_1} \delta \phi_{x_2}}, \quad \quad
        \Sigma_{x_1 x_2} \equiv 2 i \frac{\delta \Gamma_2[\phi,G]}{\delta G_{x_1 x_2}}.
\end{equation}
The 2PI self-energy is obtained from a functional derivative of $\Gamma_2$ which is represented diagrammatically
to second order in the interaction vertex in Fig.~\ref{fig:gamma2}.

The equations of motion obtained from Eqs.~(\ref{eq:eqm1}) and (\ref{eq:eqm2}) when including diagrams ``a'', ``b'', and ``c''
in $\Gamma_2$ can have as many as seven time integrals.  Hence we make several approximations to obtain
equations that are more amenable to numerical solution. First, we make the Hartree-Fock Bogoliubov (HFB)
approximation \cite{Fitzpatrick2018b} and keep only diagram ``a'' in $\Gamma_2$.  Second, we make a low-frequency
approximation.  This allows the vertices $u$ and $\tilde{u}$ to be replaced by constants $u_1$, $u_2$, and $v_1$,
details of which are given in Refs.~\cite{Fitzpatrick2018a,Fitzpatrick2018b}.  Terms involving $u_2$ can be shown to 
be small in comparison to those involving $u_1$, and hence after following through these simplifications, as outlined 
in Refs.~\cite{Fitzpatrick2018a,Fitzpatrick2018b} we obtain equations of motion for the spectral function
\begin{equation}
	A_{\bvec{k}}(t,t^\prime) = \left<\hat{a}^{\vphantom{\dagger}}_\bvec{k}(t)\hat{a}^\dagger_{\bvec{k}}(t^\prime) - \hat{a}^\dagger_{\bvec{k}}(t^\prime)\hat{a}^{\vphantom{\dagger}}_\bvec{k}(t)\right>_{\hat{\rho}_i},
\end{equation}
and the kinetic Green's function
\begin{equation}
        G_{\bvec{k}}^{(K)}(t,t^\prime) = -i \left<\hat{a}^{\vphantom{\dagger}}_\bvec{k}(t)\hat{a}^\dagger_{\bvec{k}}(t^\prime) + \hat{a}^\dagger_{\bvec{k}}(t^\prime)\hat{a}^{\vphantom{\dagger}}_\bvec{k}(t)\right>_{\hat{\rho}_i},
\end{equation}
where the expectation values are taken with respect to the initial state
\begin{equation}
	\hat{\rho}_i = \frac{e^{-\beta\hat{H}_{BHM}(t_i)}}{{\rm Tr}\left[e^{-\beta\hat{H}_{BHM}(t_i)}\right]}.
\end{equation}

\begin{widetext}
For a quench that leads to a final state in the Mott insulating regime and in which the system is initially
thermalized in the atomic limit ($J = 0$), the equations of motion take the form \cite{Fitzpatrick2018b}

\begin{eqnarray}
	A_{\bvec{k}}(t,t^\prime) & = & {\mathcal A}(t-t^\prime) - i \int_{t^\prime}^t dt^{\prime\prime} {\mathcal A}(t-t^{\prime\prime}) \Sigma_{\bvec{k}}^{\rm (HFB)}(t^{\prime\prime})  A_{\bvec{k}}(t^{\prime\prime},t^\prime), \\
	 G_{\bvec{k}}^{(K)}(t,t^\prime) & = & {\mathcal G}^{(K)}(t-t^\prime) - i \int_0^t dt^{\prime\prime} {\mathcal A}(t-t^{\prime\prime}) \Sigma_{\bvec{k}}^{\rm (HFB)}(t^{\prime\prime}) G^{(K)}_{\bvec{k}}(t^{\prime\prime},t^\prime)
	  + i\int_0^{t^\prime} dt^{\prime\prime} {\mathcal G}^{(K)}(t-t^{\prime\prime}) \Sigma_{\bvec{k}}^{\rm (HFB)}(t^{\prime\prime})  A_{\bvec{k}}(t^{\prime\prime},t^\prime), \nonumber \\ & & 
\end{eqnarray}
where ${\mathcal A}(t-t^\prime)$ and ${\mathcal G}^{(K)}(t-t^\prime)$ are the spectral function and the kinetic Green's function in the atomic ($J=0$) limit (in this limit 
both quantities are time translation invariant). [Specific forms for ${\mathcal A}$ and ${\mathcal G}^{(K)}$ are specified in Refs.~\cite{Fitzpatrick2018a,Fitzpatrick2018b}].   
The Hartee-Fock-Bogoliubov-like approximation for the self-energy (obtained by only keeping diagram ``a'' of $\Gamma_2$) takes the form
\begin{equation}
	\Sigma_{\bvec{k}}^{\rm (HFB)}(t) = \epsilon_{\bvec{k}}(t) + 2u_1 \left[n(t) - n_{J=0}\right],
\end{equation}
with 
\begin{eqnarray}
	\epsilon_{\bvec{k}} & = & -2J(t) \sum_{i=1}^d \cos(k_i a), \\
        n(t) & = & \frac{1}{N_s} \sum_{\bvec{k}} n_{\bvec{k}}(t), \\
	n_{\bvec{k}}(t) & = & \frac{1}{2}\left\{iG_{\bvec{k}}^{(K)}(t,t) - 1\right\},
\end{eqnarray}
and $u_1$ depends on $U$, $\mu$ and temperature \cite{Fitzpatrick2018a}.  
Knowledge of $n_{\bvec{k}}(t)$ -- the particle distibution over the quasi-momentum $\bvec{k}$ at time t -- allows us to calculate the single-particle density matrix.
The full expression for $u_1$ is
\begin{eqnarray}
u_{1} & = & -\frac{2\left\{ \mathcal{G}^{12,\left(R\right)}\left(\omega^{\prime}=0\right)\right\} ^{-4}}{\mathcal{Z}_{0}}\nonumber \\
 &  & \quad\times\sum_{n=0}^{\infty}e^{-\beta\left(\mathcal{E}_{n}-\mathcal{E}_{n_{0}}\right)}\left\{ \frac{\left(n+1\right)\left(n+2\right)}{\left(\mathcal{E}_{n+2}-\mathcal{E}_{n}\right)\left(\mathcal{E}_{n+1}-\mathcal{E}_{n}\right)^{2}}+\frac{n\left(n-1\right)}{\left(\mathcal{E}_{n-2}-\mathcal{E}_{n}\right)\left(\mathcal{E}_{n-1}-\mathcal{E}_{n}\right)^{2}}\right.\nonumber \\
 &  & \left.\phantom{\quad\times\sum_{n=0}^{\infty}e^{-\beta\left(\mathcal{E}_{n}-\mathcal{E}_{n_{0}}\right)}}\quad-\frac{\left(n+1\right)^{2}}{\left(\mathcal{E}_{n+1}-\mathcal{E}_{n}\right)^{3}}-\frac{n^{2}}{\left(\mathcal{E}_{n-1}-\mathcal{E}_{n}\right)^{3}}\right.\nonumber \\
 &  & \left.\phantom{\quad\times\sum_{n=0}^{\infty}e^{-\beta\left(\mathcal{E}_{n}-\mathcal{E}_{n_{0}}\right)}}\quad-\frac{n\left(n+1\right)}{\left(\mathcal{E}_{n+1}-\mathcal{E}_{n}\right)\left(\mathcal{E}_{n-1}-\mathcal{E}_{n}\right)^{2}}-\frac{n\left(n+1\right)}{\left(\mathcal{E}_{n+1}-\mathcal{E}_{n}\right)^{2}\left(\mathcal{E}_{n-1}-\mathcal{E}_{n}\right)}\right\} ,\label{eq:u_1 defined - 1}
\end{eqnarray}
where
\begin{eqnarray}
\mathcal{G}^{12,\left(R\right)}\left(\omega^{\prime}=0\right) & = & -\frac{1}{\mathcal{Z}_{0}}\sum_{n=0}^{\infty}e^{-\beta\left(\mathcal{E}_{n}-\mathcal{E}_{n_{0}}\right)}\left\{ \frac{\left(n+1\right)}{\mathcal{E}_{n+1}-\mathcal{E}_{n}}+\frac{n}{\mathcal{E}_{n-1}-\mathcal{E}_{n}}\right\} , \label{eq:static limit of G0^(R) - 1}
\end{eqnarray}
\end{widetext}
and  $\mathcal{Z}_{0}$ is the atomic partition function
\begin{eqnarray}
\mathcal{Z}_{0} & \equiv & \sum_{n=0}^{\infty}e^{-\beta\left(\mathcal{E}_{n}-\mathcal{E}_{n_{0}}\right)},\label{eq:atomic partition function - 1}
\end{eqnarray}
with $n_0 =  \left\lceil \mu/U\right\rceil$ and
\begin{equation}
	\mathcal{E}_n = \frac{U}{2}n(n-1) - n\mu.
\end{equation}

\begin{widetext}

\begin{figure}[h]
    \includegraphics[width=6cm]{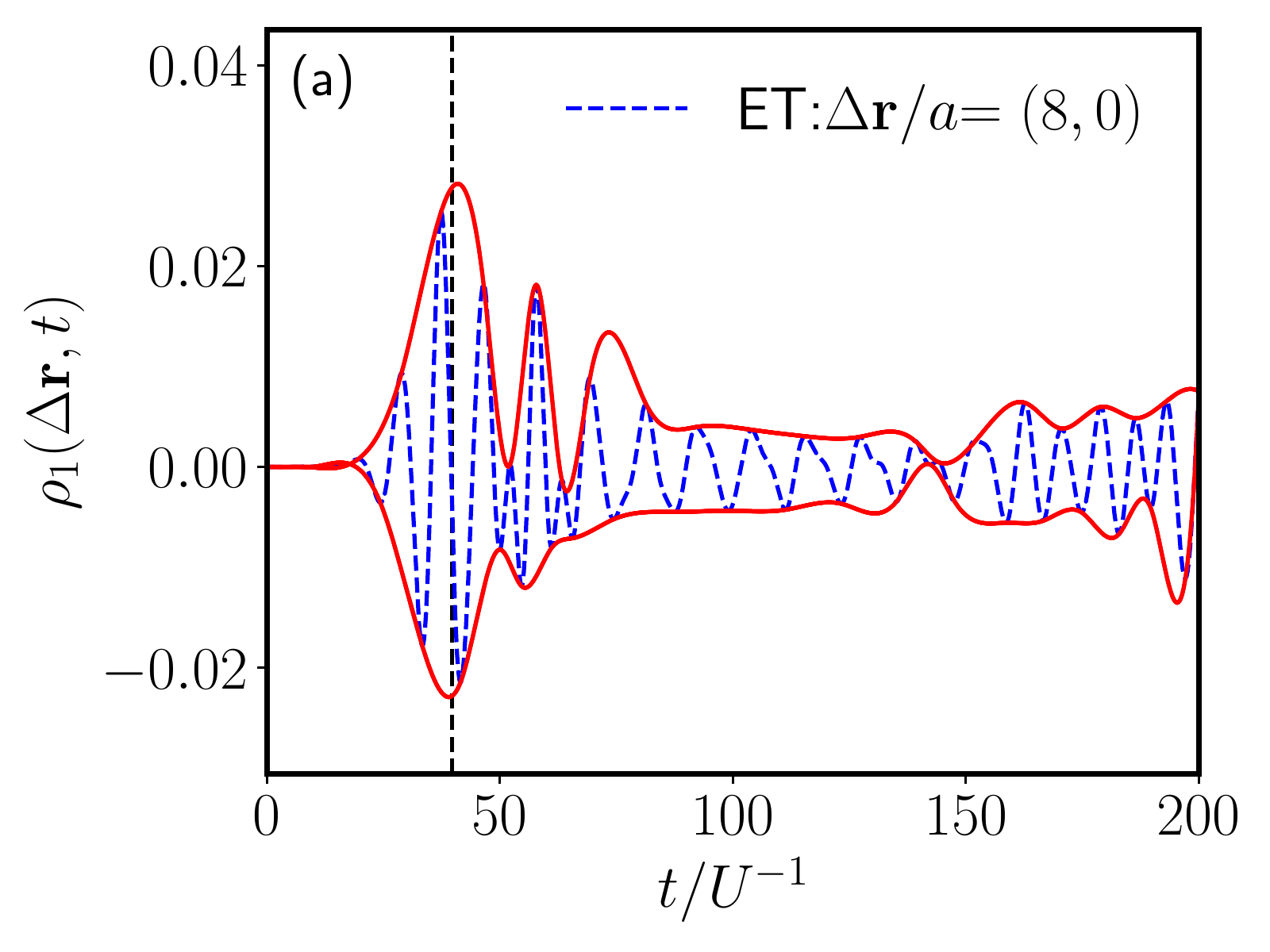} 
    \includegraphics[width=6cm]{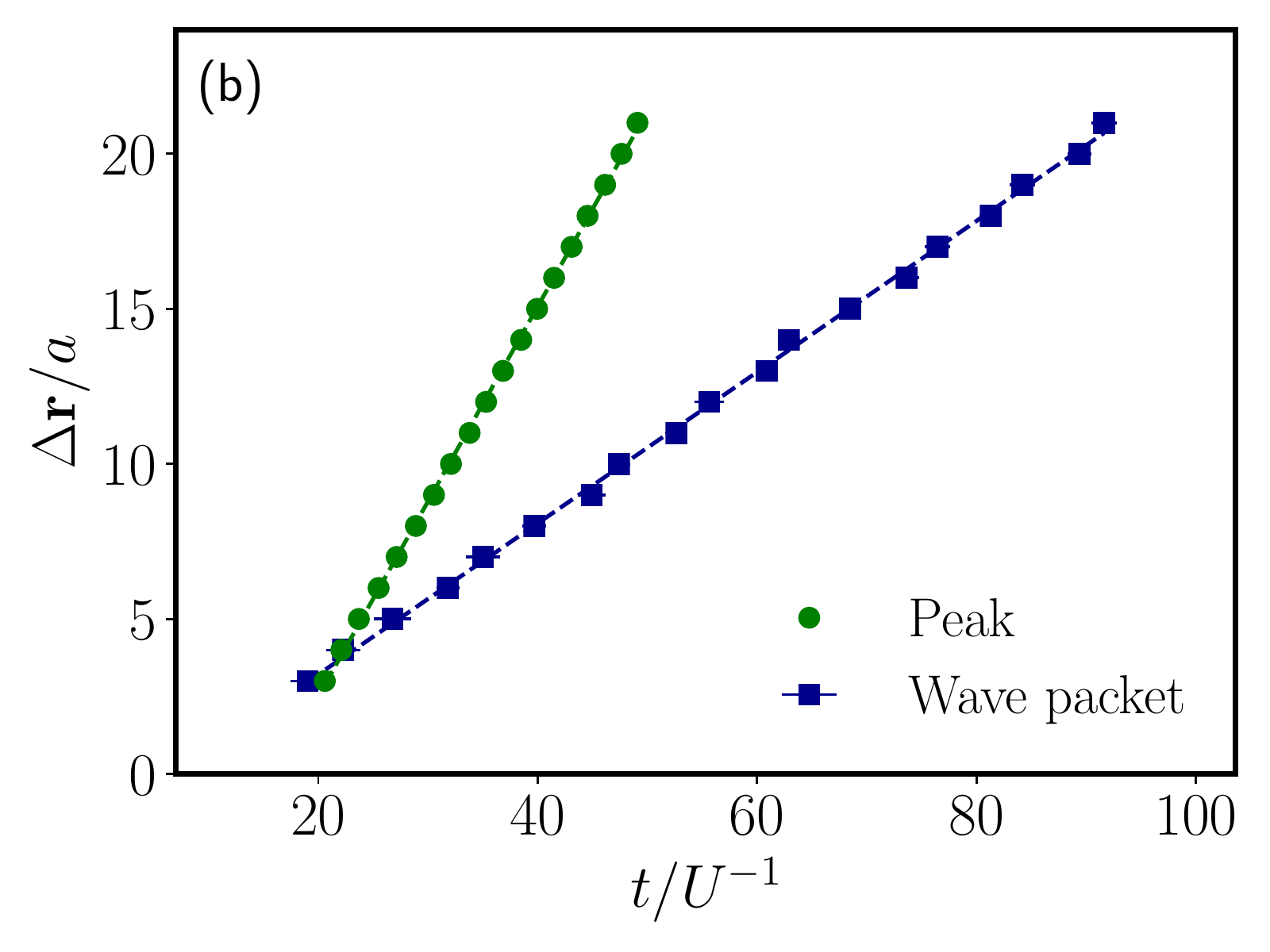} \\
    \includegraphics[width=6cm]{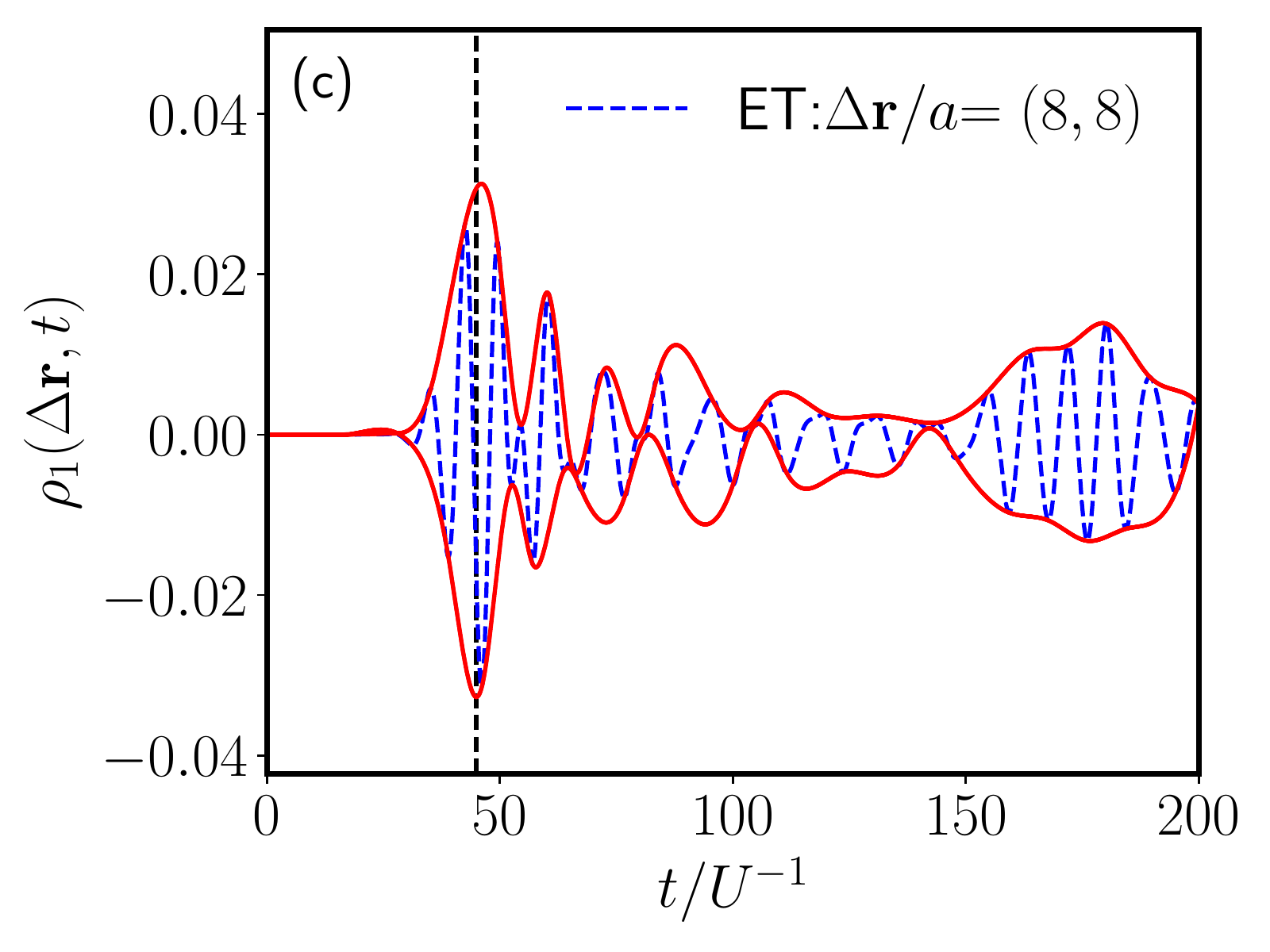}
    \includegraphics[width=6cm]{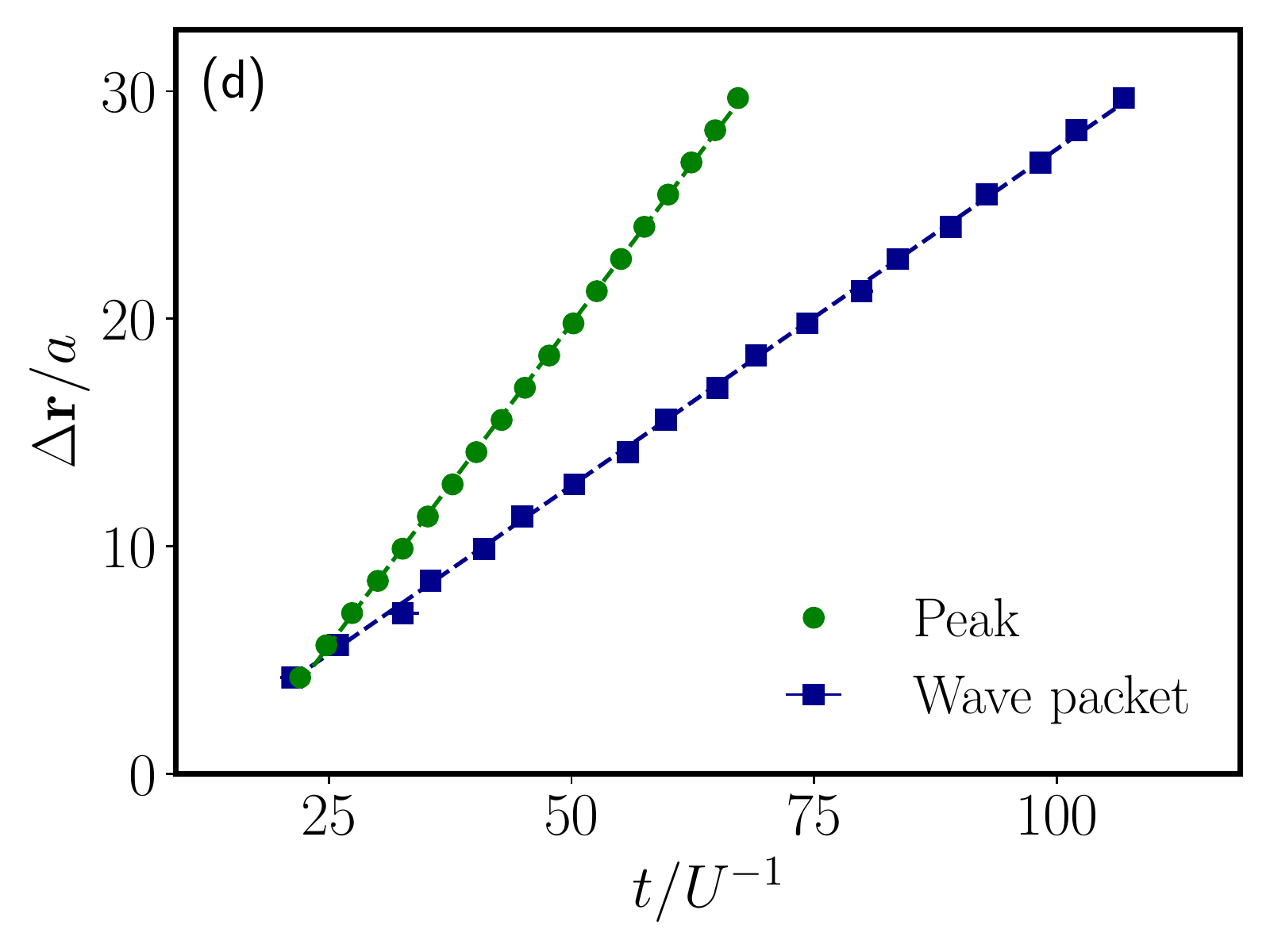}
\caption{ Tracking the wavefront for a  $50 \times 50$ system. (a) and (c): Dynamics of $\rho_1(\Delta {\bvec{r}},t)$ for $\Delta {\bvec{r}}/a= (8,0)$ and
        $\Delta {\bvec{r}}/a= (8,8)$; (b) and (d): scatterplots of the times $t/U^{-1}$ for the maximum peak (green) and the wave packet (blue) to travel
        a particle separation distance $\Delta r/a$ along the crystal axes and diagonals respectively.
        In (a) and (c) the solid red lines trace the envelopes of the wave packets, while the dashed black vertical lines indicate the positions of the wave packets.
         In (b) and (d) we fit straight lines to the data  to obtain the group and phase velocities along the crystal axes and diagonals.
         The parameters are $\beta U = 1000$, $U/J_f = 28.6$, $\mu/U = 0.4116$, $t_c /U^{-1} = 5$, $t_Q/ U^{-1} = 0.1$.}
\label{fig:2d_scatterplots}
\end{figure}
\end{widetext}

The Hartree-Fock-Bogoliubov self-energy is most accurate in the strongly interacting limit and is least accurate at the tips of the Mott
lobes.  This is illustrated in Fig.~6 of Ref.~\cite{Fitzpatrick2018a}, where phase boundaries between the Mott and superfluid phases were
calculated with the HFB self-energy and compared to both mean-field theory and exact numerical results. In dimensions 1, 2 and 3, the HFB 
self-energy leads to a significant improvement beyond mean-field theory and in dimensions 2 and 3 it is only at
the tips of the Mott lobes where there is some deviation between exact and HFB results.  The deviations in 1 dimension are larger, but even there, 
the HFB result is a dramatic improvement on mean-field theory.

The explicit form of $J(t)$ that we consider is
\begin{equation}
        J(t) = \frac{J_f}{2} \left[1 + \tanh\left(\frac{t-t_c}{\tau_Q}\right)\right],
\end{equation}
where $J_f$ characterizes the final hopping strength, $t_c$ characterizes the time at which the middle of the quench occurs, and $\tau_Q$
characterizes the duration of the quench.
The equations of motion do not have any known analytical solution, and hence we have solved
them numerically using a block-by-block scheme detailed in Ref.~\cite{Fitzpatrick2018b}.  We give additional details about the solutions of
these equations in the following section.

\begin{widetext}

\section{Single-particle correlations}
	\label{sec:appB}
In this appendix we provide additional details relating to the methodology we used to identify the group velocity and
phase velocity for the spreading of correlations from our calculations of $\rho_1(\Delta {\bvec{r}},t)$.
We discussed the one-dimensional case in the main body 
of the paper, but provide additional details for the two- and three-dimensional cases here.  
For a given $U/J_f$, we calculate $\rho_1(\Delta\bvec{r},t)$ 
and for each value of $\Delta \mathbf{r}$ we obtain the timewise positions of the wave packet, 
and the largest peak [i.e. the point in time where $\rho_1\left(\Delta\mathbf{r}, t\right)$ takes its maximum value].

\begin{figure}[th]
    \includegraphics[width=5cm]{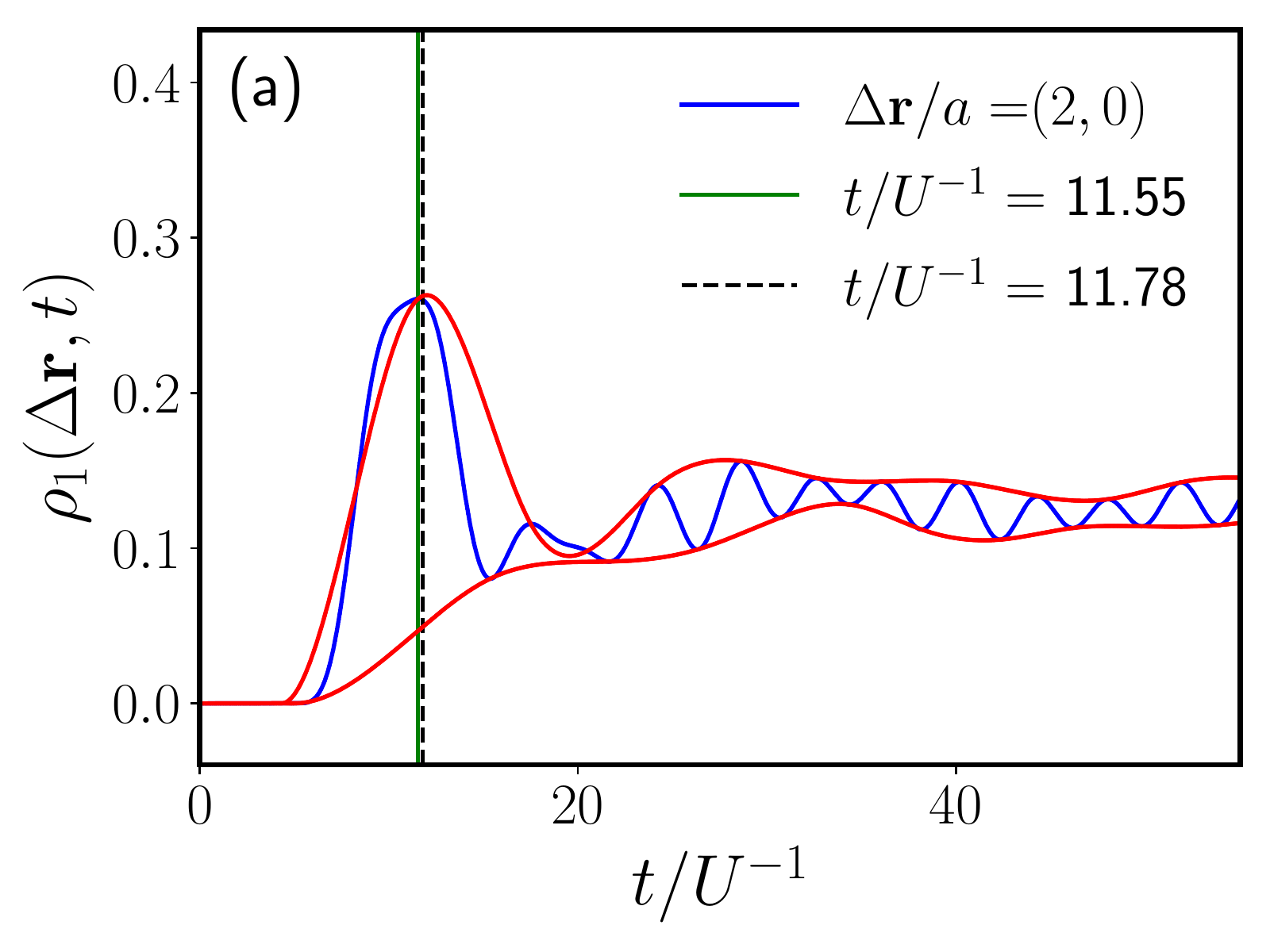}
    \includegraphics[width=5cm]{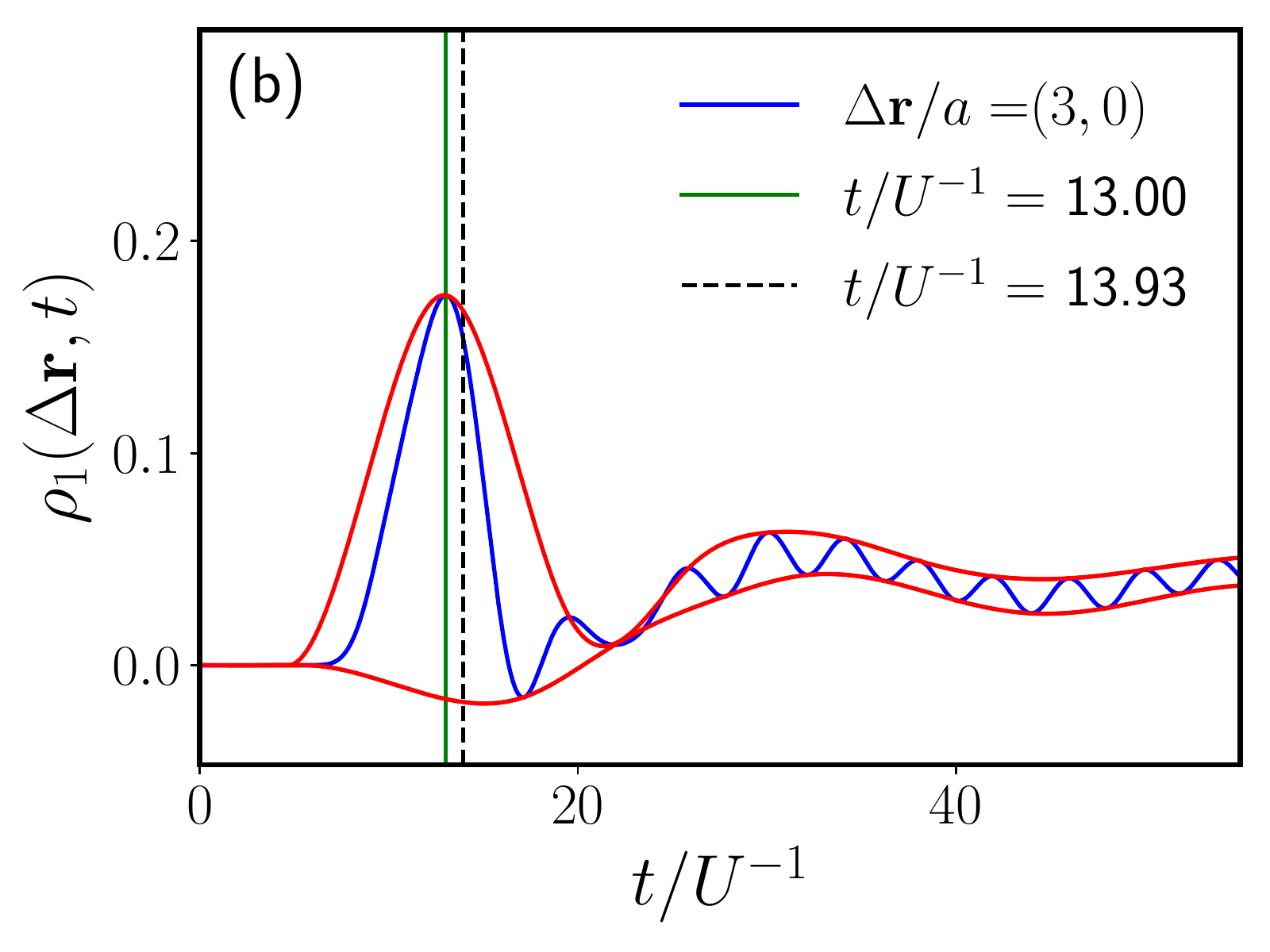}
    \includegraphics[width=5cm]{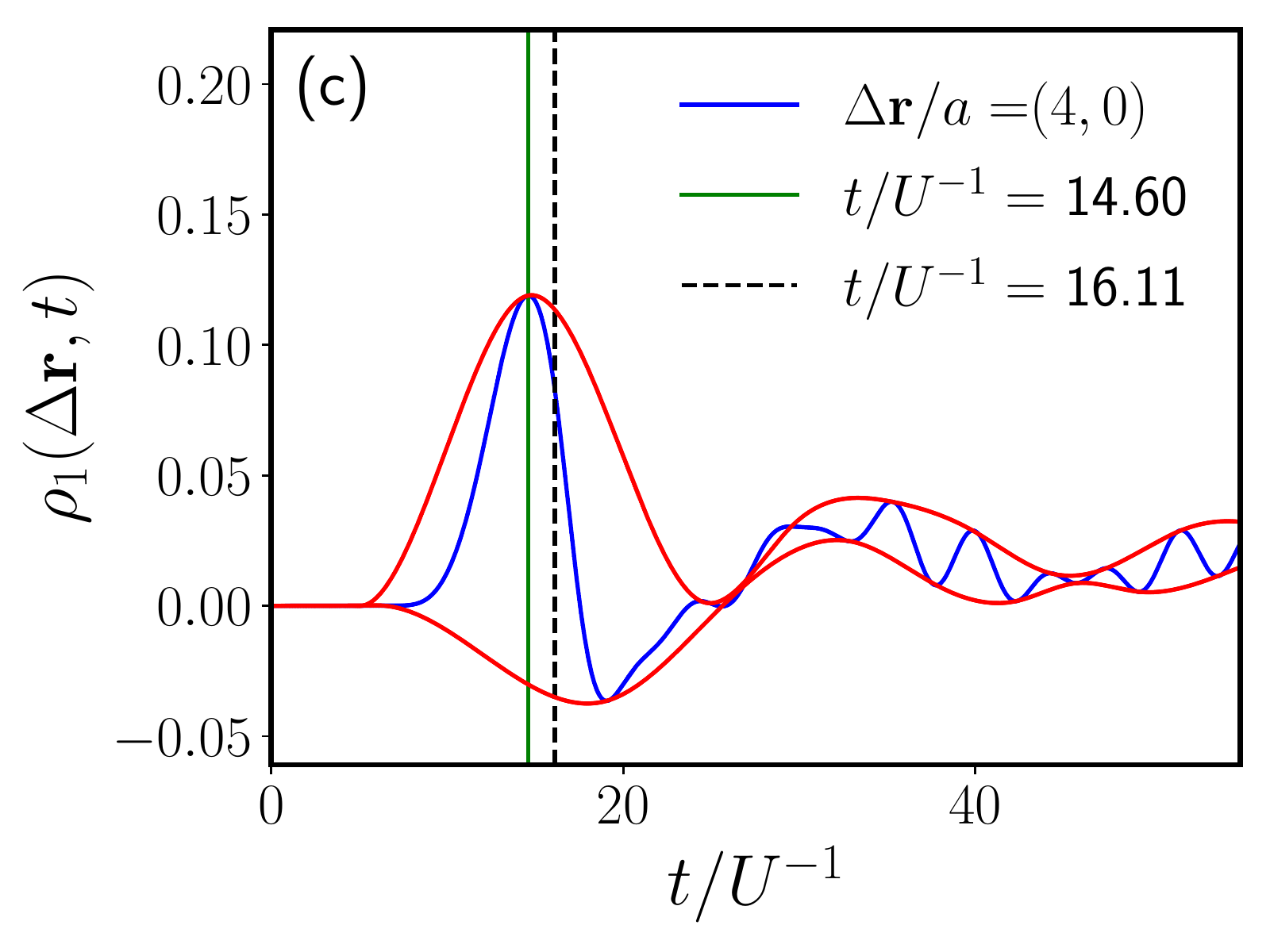}
    
    \includegraphics[width=5cm]{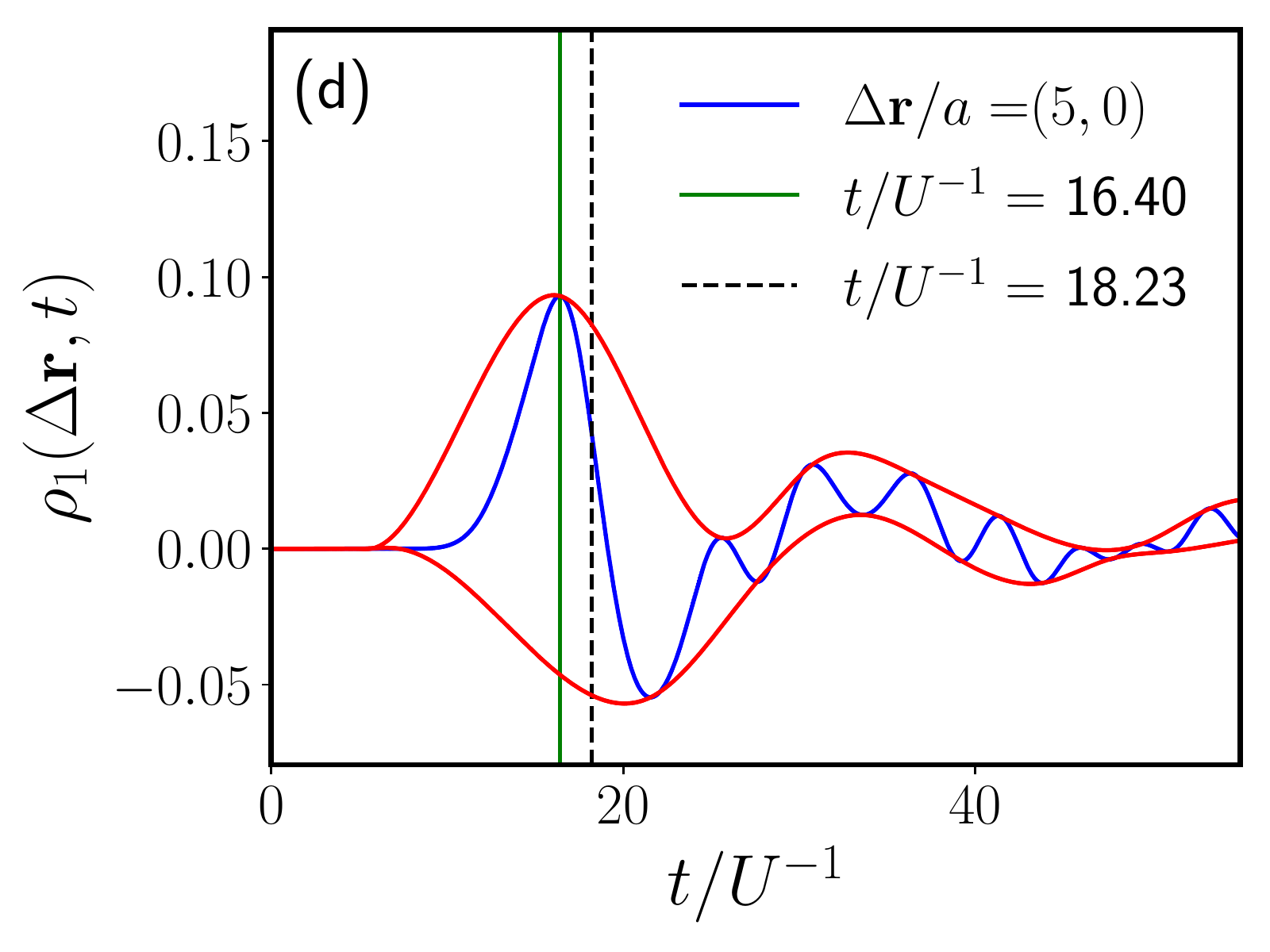}
    \includegraphics[width=5cm]{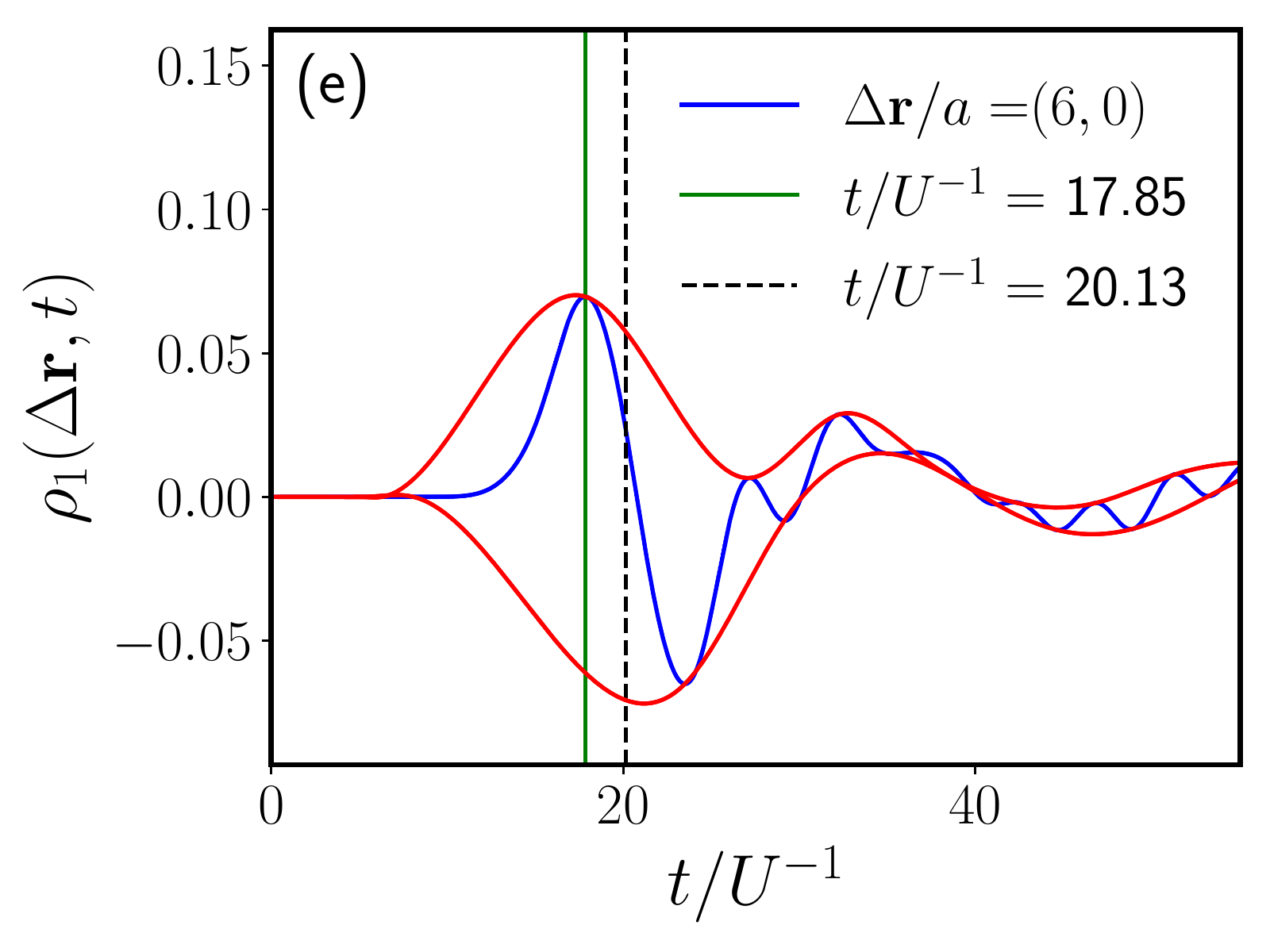}
        \caption{(a)-(e) Dynamics of $\rho_1(\Delta {\bvec{r}},t)$ along the crystal axis from $\Delta {\bvec{r}}/a= (2,0)$ to $\Delta {\bvec{r}}/a= (6,0)$.
        The solid vertical green lines show the positions of the peaks related to the phase velocity and
        dashed vertical black lines show the positions of the wave packets related to group velocity.
        The parameters in (a)–(e) are $\beta U = 1000$, $U/J_f = 19.6$, $\mu/U = 0.4116$, $t_c /U^{-1} = 5$, $t_Q/ U^{-1} = 0.1$, and the system size was 50$\times$50.}
\label{fig:2d_crystal_axis_phase_group}
\end{figure}

We  track the propagation of the maximum peak of the $\rho_1\left(\Delta\mathbf{r}, t\right)$ time series to extract the phase velocity.
For the group velocity we first locate all local maxima and minima of $\rho_1\left(\Delta \mathbf{r}, t\right)$ and then use these points to construct
upper and lower envelopes using both linear and cubic interpolation.  Having found two different pairs of upper and lower
envelopes, one from the linear interpolation and one from the cubic interpolation, we find the times $t_{\rm max}$ and $t_{\rm min}$ where each envelope
has a maximum or minimum respectively.  To find the centre of the wavepacket we average $t_{\rm max}$ and $t_{\rm min}$ in three different ways, for each pair of upper and lower
envelopes (i.e. we consider 6 different estimates for the centre of the wavepacket).

The averages we consider are
\begin{equation}
	C_1 = \frac{t_{\rm max} + t_{\rm min}}{2},
\end{equation}
and
\begin{equation}
	C_2 = \frac{\alpha t_{\rm max} + \beta t_{\rm min}}{\alpha + \beta},
\end{equation}
where $\alpha = \left|\rho_{\rm max} - \rho_{\rm av}\right|$ and $\beta = \left|\rho_{\rm av} - \rho_{\rm min}\right|$,
with $\rho_{\rm max}$ the maximum value of $\rho_1\left(\Delta\mathbf{r}, t\right)$, $\rho_{\rm min}$ the minimum 
value of $\rho_1\left(\Delta\mathbf{r}, t\right)$ and $\rho_{\rm av}$ the average of $\rho_1\left(\Delta\mathbf{r}, t\right)$.
We calculate $C_2$ with i) $\rho_{\rm av}$ averaged over the whole time interval and ii) with $\rho_{\rm av}$ averaged
over the time interval $[0, t_{\rm max} + 5U^{-1}]$ (to capture the average of  $\rho_1\left(\Delta\mathbf{r}, t\right)$ 
in the only the first peak).  We then average $C_1$ and the two different versions of $C_2$
for both sets of envelopes to determine the centre of the envelope.  These multiple approaches allow us to estimate the 
uncertainty in the centre of the envelope, which is generally small, e.g. see Fig.~\ref{fig:2d_scatterplots} b) and d).
By tracking the propagation of the wave packet, we can extract the group velocity for the spreading of 
single-particle correlations \cite{Fitzpatrick2018b}. 

In Fig.~\ref{fig:2d_scatterplots} (a) and (c) we display the time evolution of $\rho_1(\Delta {\bvec{r}},t)$ for $\Delta \bvec{r}/a =(8,0)$ 
 (i.e., along the crystal axis) and $\Delta \bvec{r}/a =(8,8)$ (i.e., along the diagonal) respectively. Figures~\ref{fig:2d_scatterplots} (b) and (d)
plot the times $t/U^{-1}$ for the maximum peak and the wave packet to travel a particle separation distance $\Delta r/a$ along 
the crystal axes (b) and diagonals (d) respectively.
 The calculations are for $U/J_f =28.6$ and a $50 \times 50$ lattice.

\begin{figure}[h]
    \includegraphics[width=16cm]{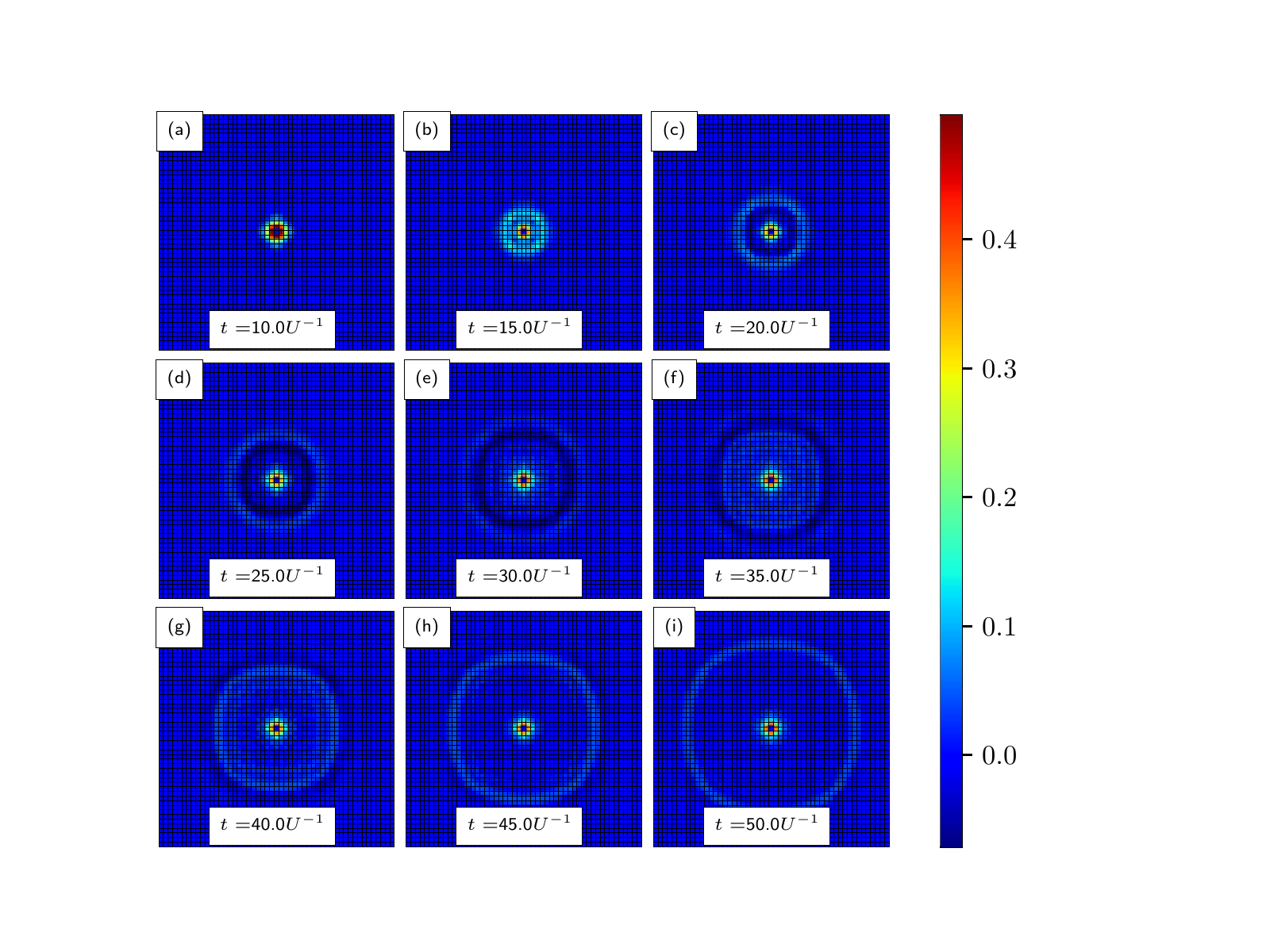}
        \caption{(a)-(i) Spatial dependency of $\rho_1\left(\Delta \mathbf{r}, t\right)$ at different moments in time $t/U^{-1}$ for a $50 \times 50$ site system.
        The parameters used are $\beta U = 1000$, $U/J_f = 19.6$, $\mu/U = 0.4116$, $t_c /U^{-1} = 5$, $t_Q/ U^{-1} = 0.1$.}
\label{fig:2d_spreading}
\end{figure}

Similarly to the one-dimensional case, we calculate  $\rho_1(\Delta {\bvec{r}},t)$ for a variety of $U/J_f$ in the Mott phase and calculate the phase and group velocities 
for correlation spreading.  To illustrate the process of determining the phase and group velocities more clearly, in Fig.~\ref{fig:2d_crystal_axis_phase_group} we show snapshots of  $\rho_1(\Delta {\bvec{r}},t)$
from ${\bvec{r}}/a= (2,0)$ to ${\bvec{r}}/a= (6,0)$ along the crystal axis. 
For each plot, we mark the timewise positions of the maximum peak and the wave packet by vertical solid green and dashed vertical black lines respectively.
The increasing gap between the two lines with increasing distance
along the crystal axis illustrates the difference between the group and phase velocities.

In Ref.~\cite{Takasu2020} the authors used both peaks and troughs to determine the phase velocity of excitations.  In Fig.~\ref{fig:2d_spreading}
we illustrate the spreading of correlations in two dimensions for $U/J_f = 19.6$.  Peaks are visible as light blue circles and troughs are visible as dark blue circles.
There is some anisotropy in the spreading, but it is relatively small since the value of $U/J_f$ is close to the critical value.

\begin{figure}[h]
    \includegraphics[width=7cm]{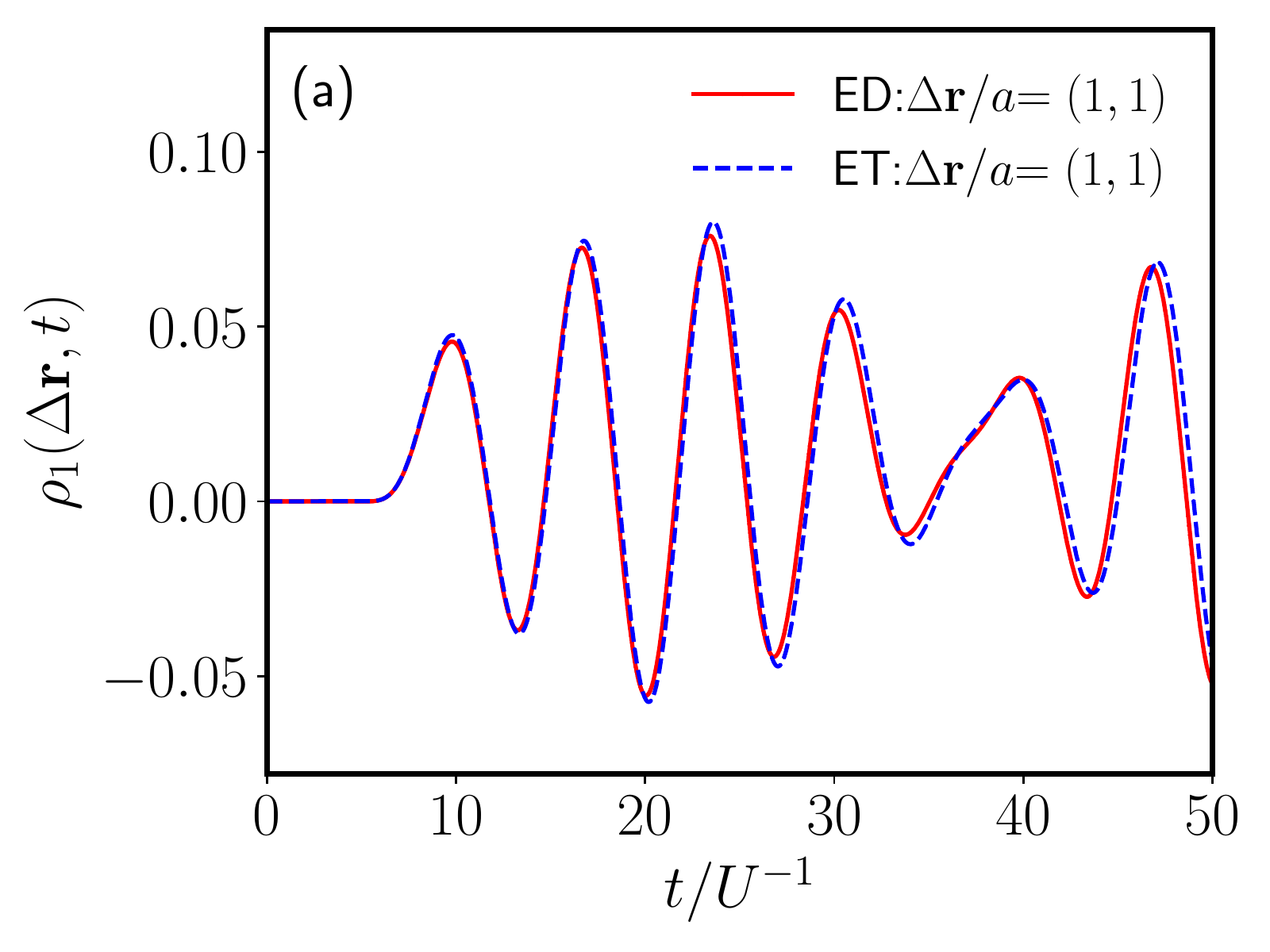}
    \includegraphics[width=7cm]{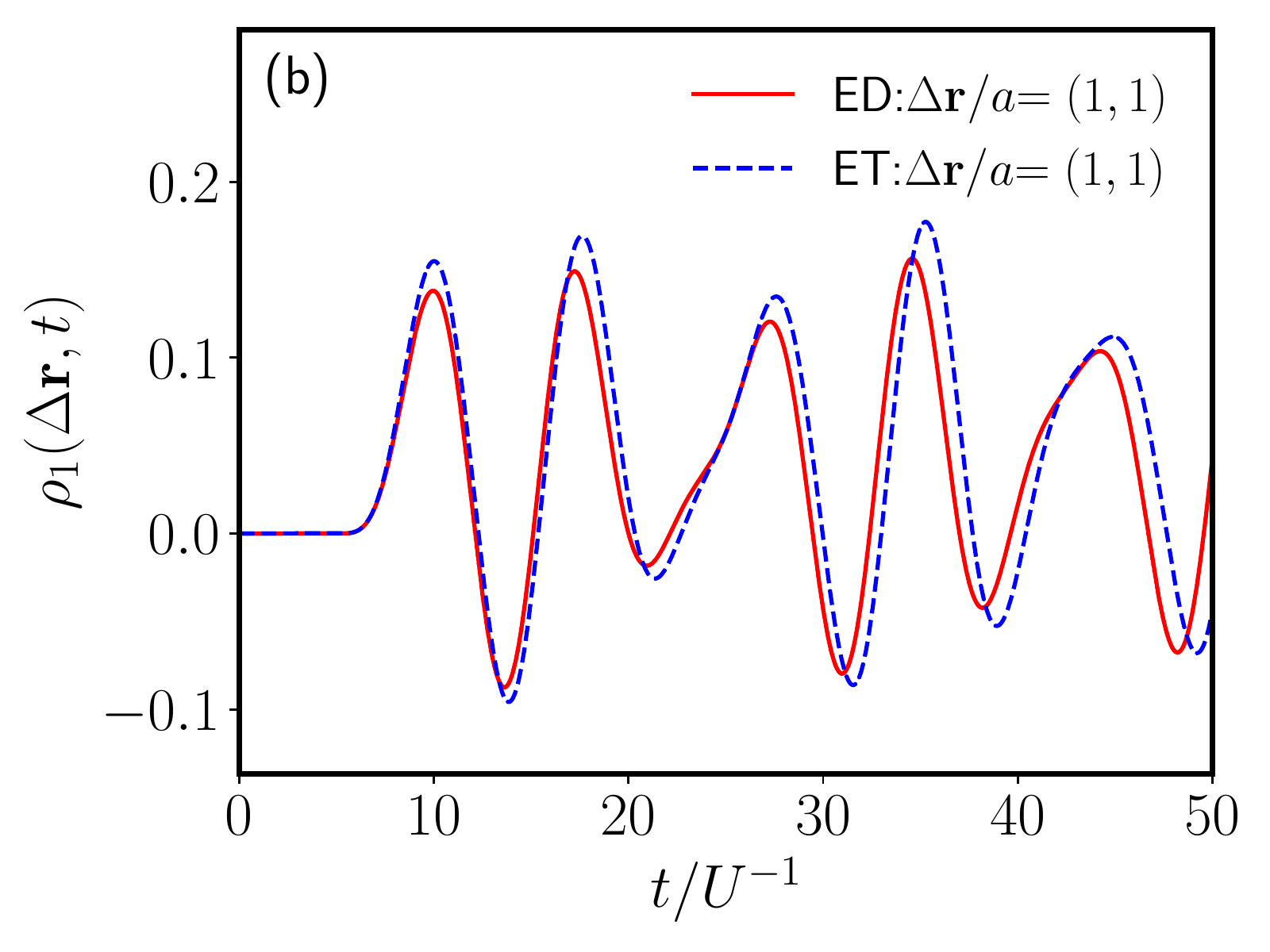} \\
    \includegraphics[width=7cm]{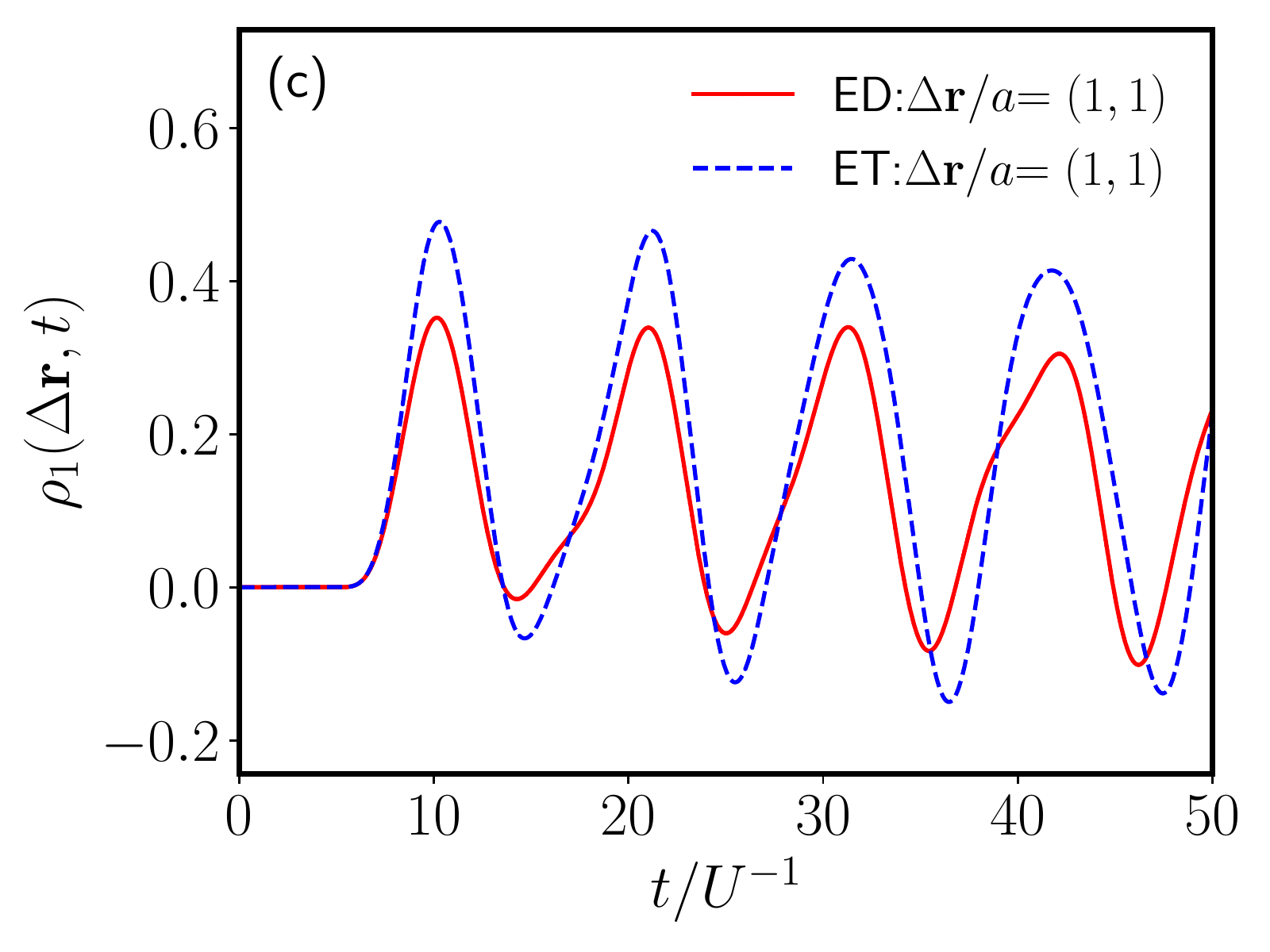}
    \includegraphics[width=7cm]{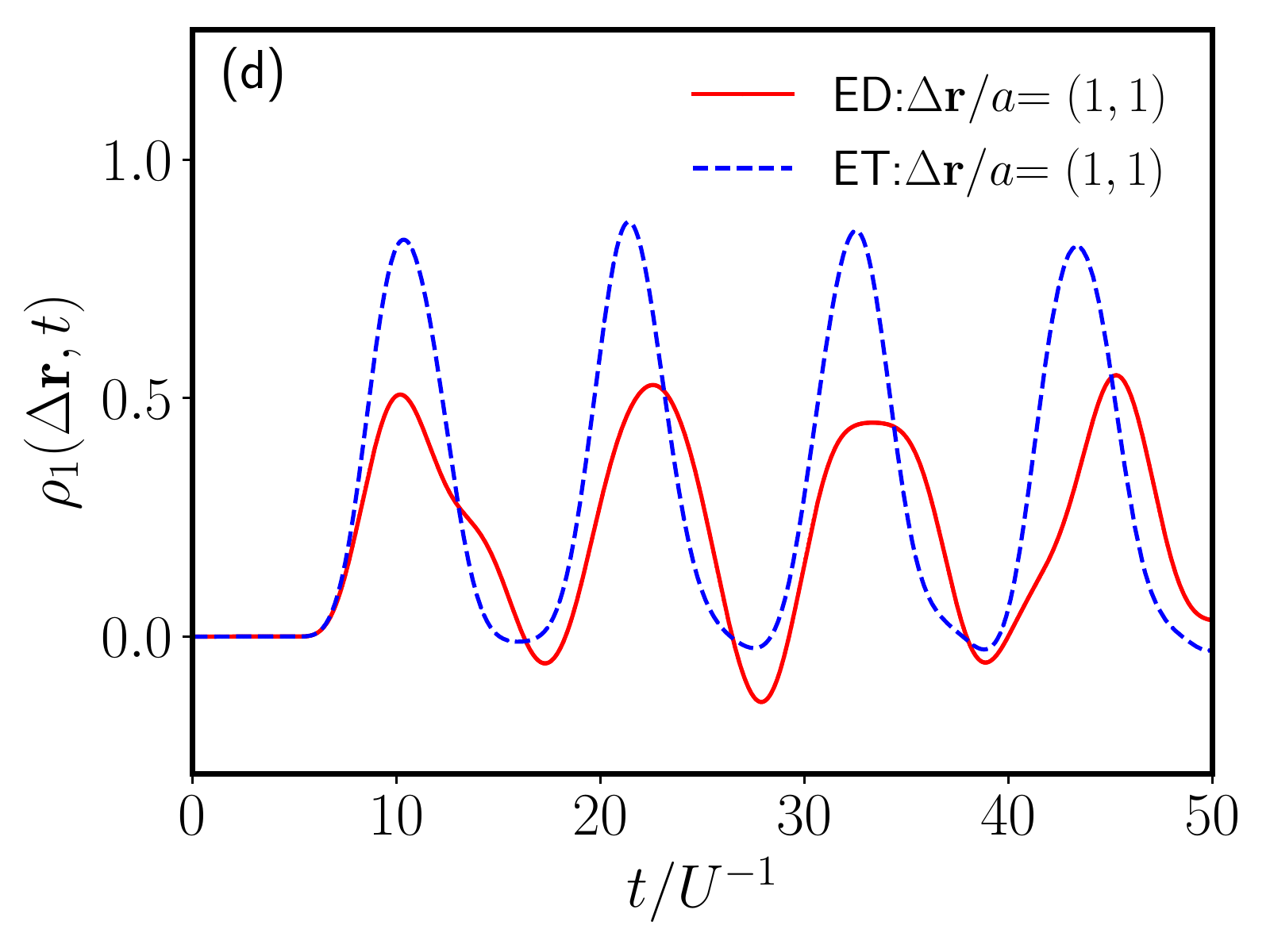}
	\caption{Comparison of exact diagonalization (ED) calculations and effective theory (ET) calculations of $\rho_1(\Delta {\bvec{r}},t)$ for $\Delta \bvec{r}/a =(1,1)$
        for interaction strengths: a) $U/J_f$ = 66.7, b) $U/J_f = 40$, c) $U/J_f = 25$ and $U/J_f = 19.6$.
         The parameters are $\beta U = 1000$, $\mu/U = 0.4116$, $t_c /U^{-1} = 5$, $t_Q/ U^{-1} = 0.1$.}
\label{fig:2d_EDvsET}
\end{figure}

As a test of our approach in two dimensions, we performed exact diagonalization calculations on a $3\times 3$ system, and compared
the results for $\rho_1(\Delta {\bvec{r}},t)$ to the results of our effective theory.  The comparisons between both methods are shown in Fig.~\ref{fig:2d_EDvsET}.  
We find that the quantitative agreement 
between the effective theory and ED is excellent at large values of $U/J_f$, but becomes less accurate for values of $U/J_f$ close
to the transition, where there is a discrepancy in the magnitude of $\rho_1(\Delta {\bvec{r}},t)$ by roughly a factor of 2.  However,
the phase of  $\rho_1(\Delta {\bvec{r}},t)$, in particular, the position of the first peak, is represented accurately by the 
effective theory.

We also give a sample of some of the calculations of  $\rho_1(\Delta {\bvec{r}},t)$ we calculated for three dimensions, showing traces of  $\rho_1(\Delta {\bvec{r}},t)$
along three different crystal directions in Fig.~\ref{fig:3d_envelopes} and fits used to determine the group and phase velocities along the respective directions 
for $U/J_f = 55.6$ in Fig.~\ref{fig:3d_scatterplots}.

\begin{figure}[ht]
    \includegraphics[width=5cm]{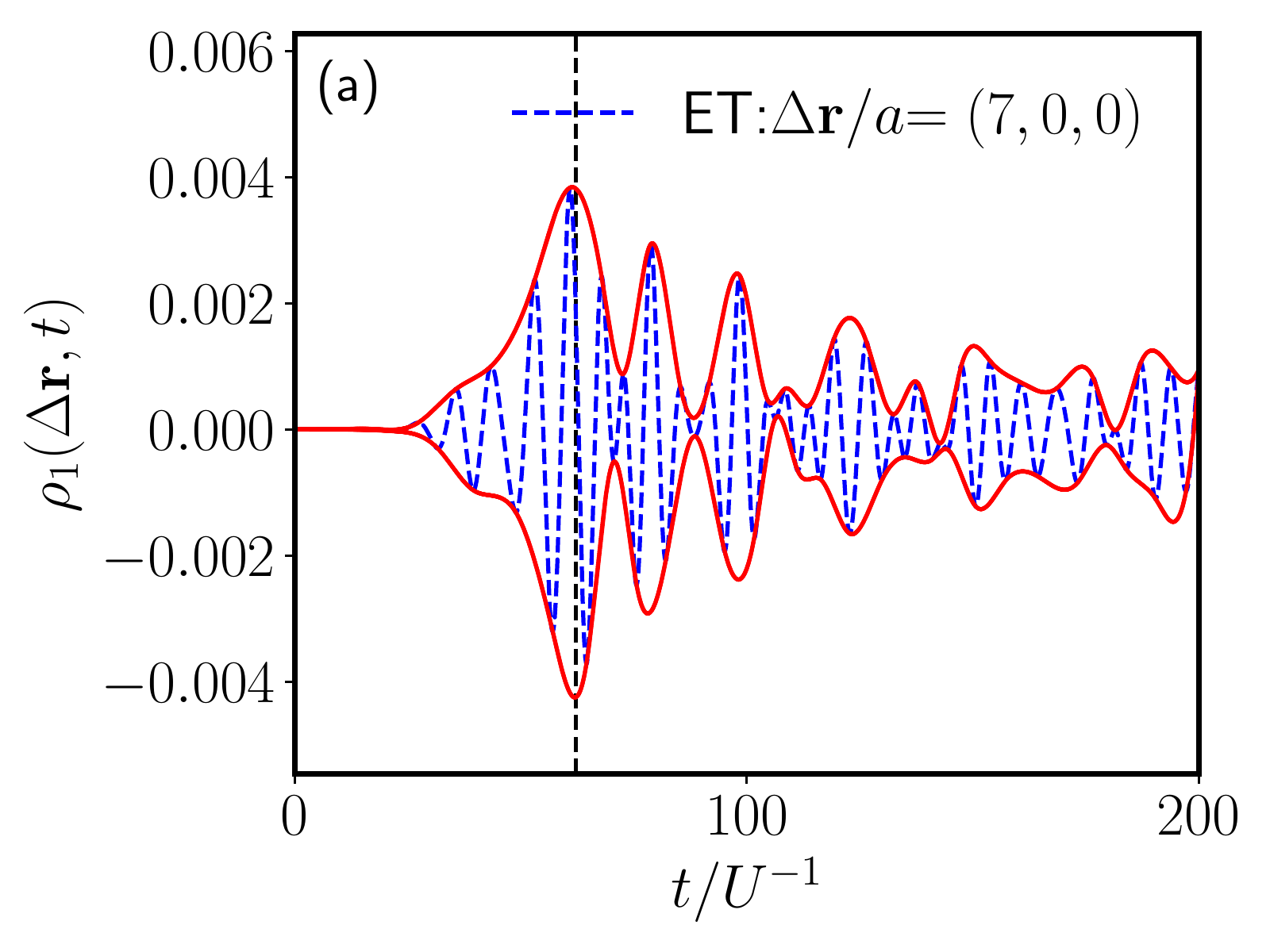}
    \includegraphics[width=5cm]{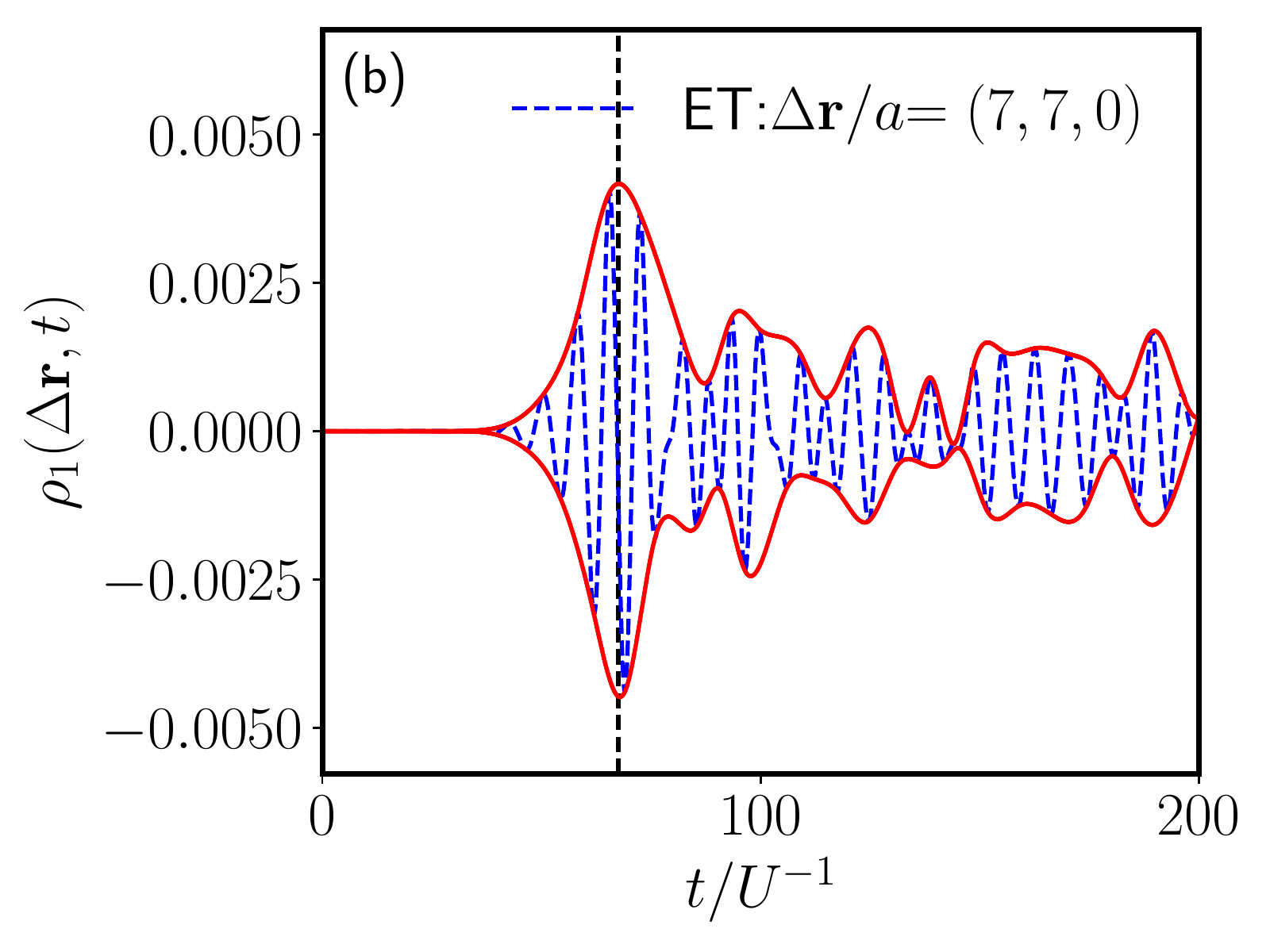}
    \includegraphics[width=5cm]{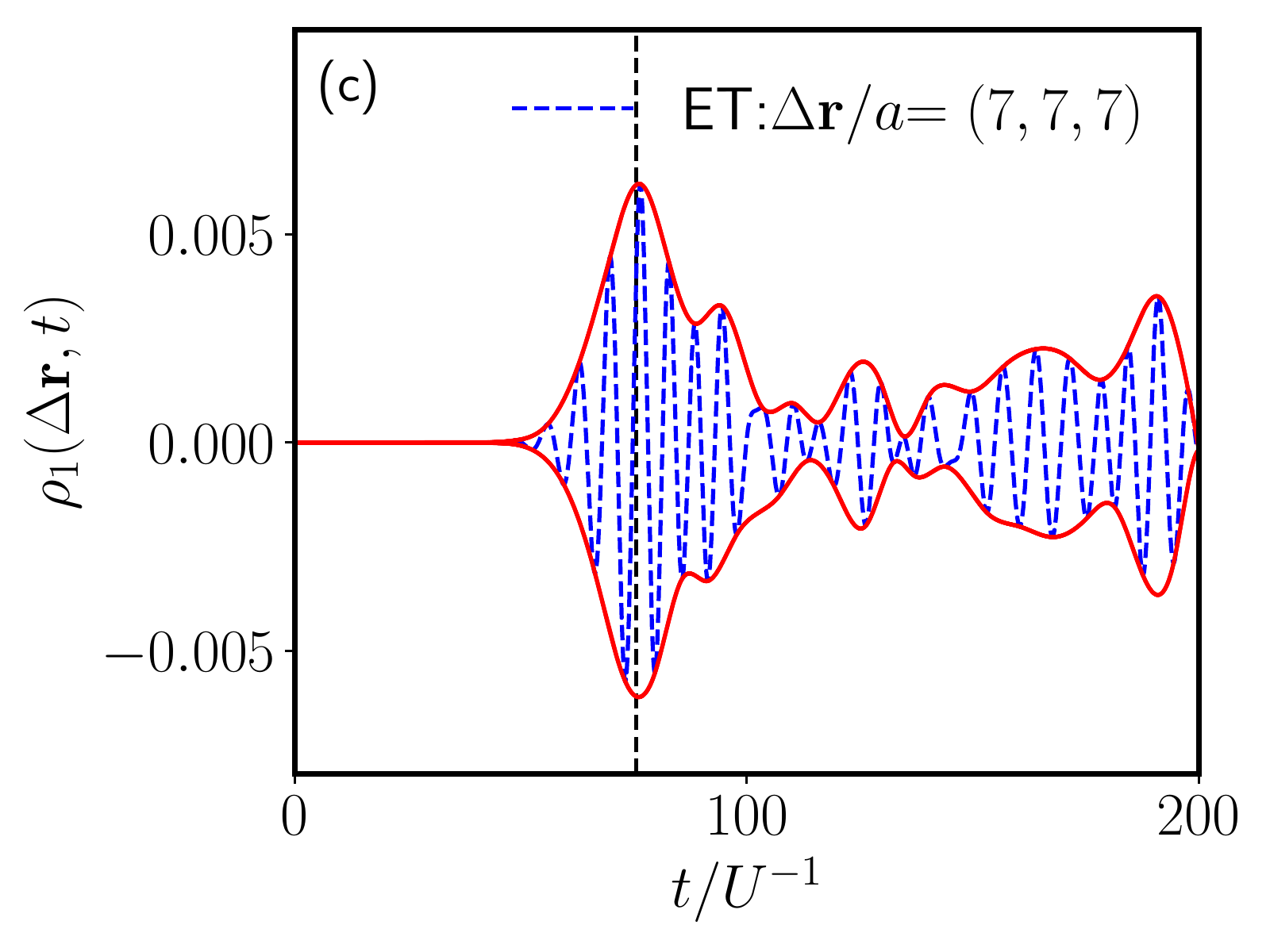}
	\caption{ Tracking the wavefront in a  $28 \times 28 \times 28$ system. Dynamics of $\rho_1(\Delta {\bvec{r}},t)$ for (a) $\Delta {\bvec{r}}/a= (7,0,0)$, (b)  $\Delta {\bvec{r}}/a= (7,7,0)$
	and (c)  $\Delta {\bvec{r}}/a= (7,7,7)$.
	The solid red
lines trace the envelopes of the wave packets, while the dashed black vertical lines indicate the positions of the wave packets.
         The parameters are $\beta U = 1000$, $U/J_f = 55.6$, $\mu/U = 0.4116$, $t_c /U^{-1} = 5$, $t_Q/ U^{-1} = 0.1$.}
\label{fig:3d_envelopes}
\end{figure}

\begin{figure}[ht]
    \includegraphics[width=5cm]{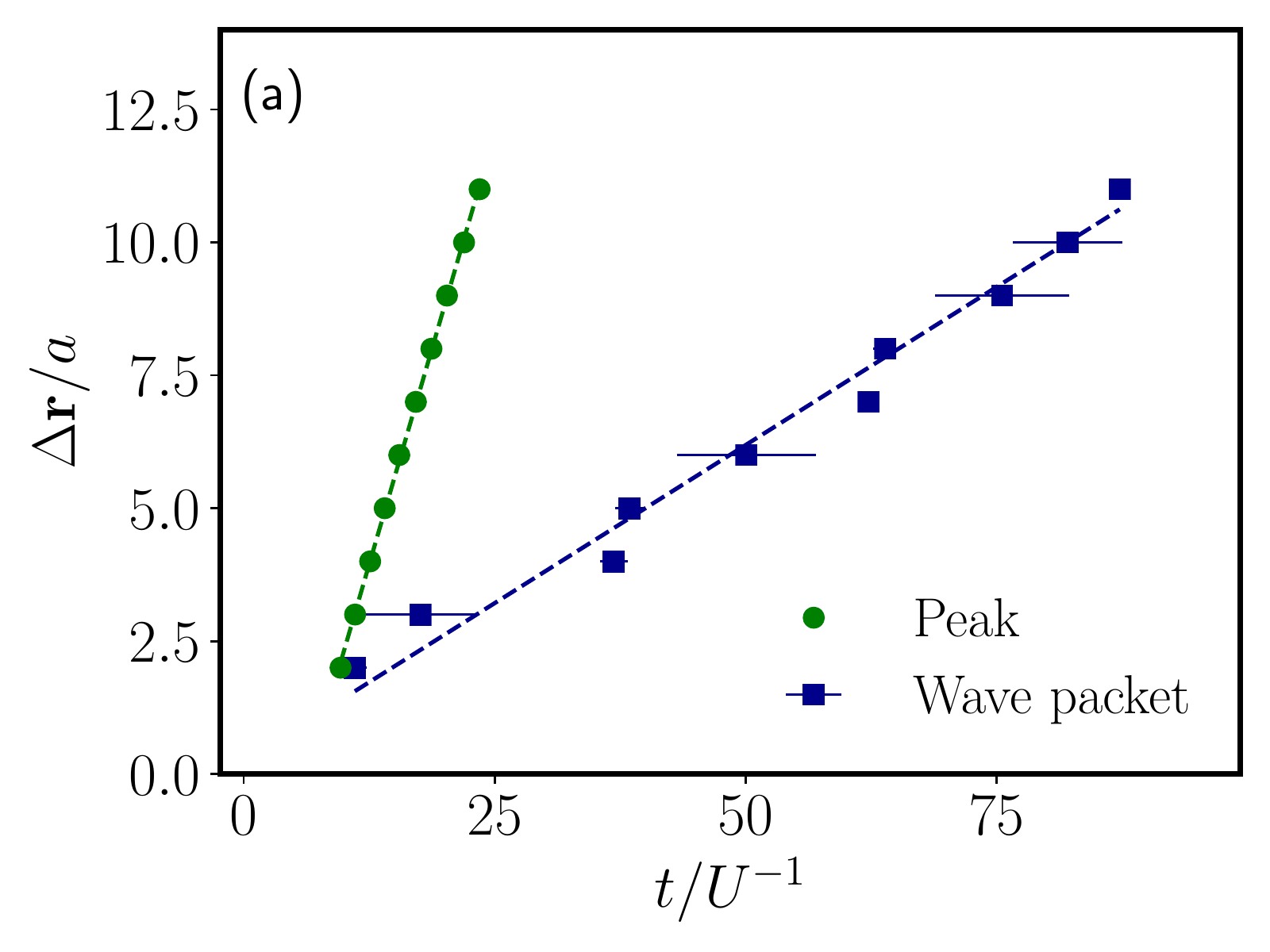}
    \includegraphics[width=5cm]{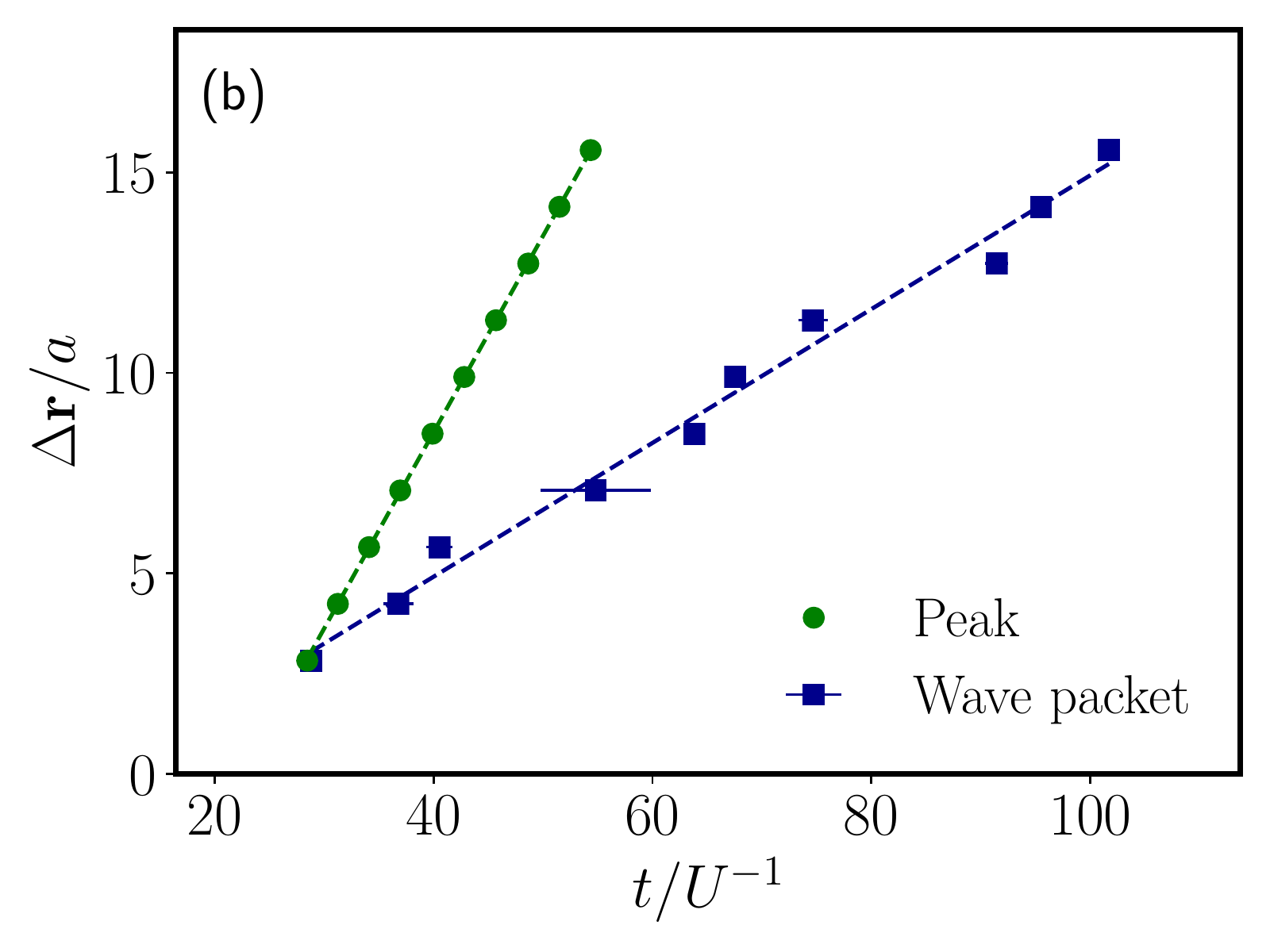}
    \includegraphics[width=5cm]{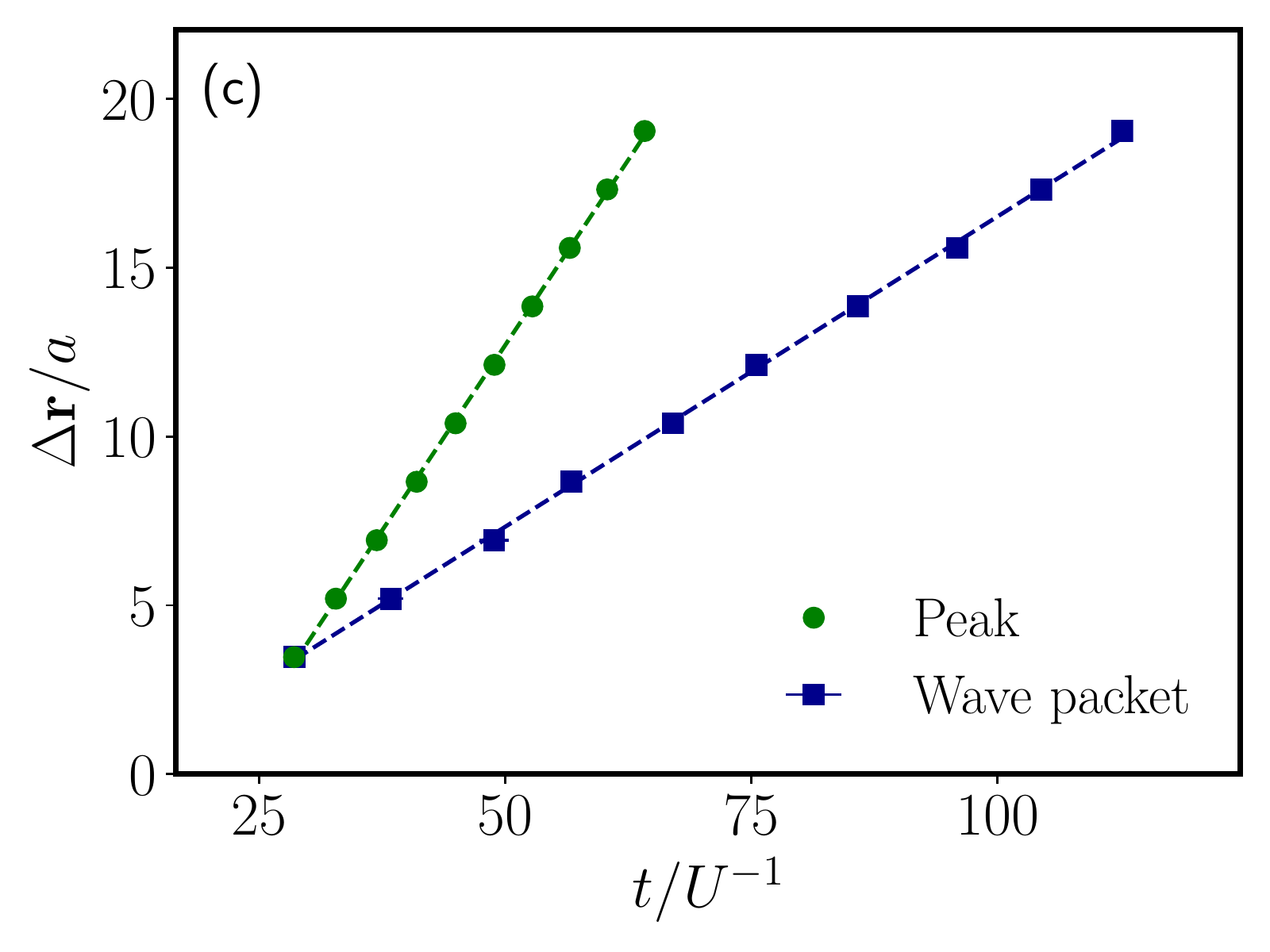}
	\caption{Scatterplots of the time $t/U^{-1}$ for the maximum peak (green) and the wave packet (blue) to travel a distance $\Delta r/a$ 
	 along (a) the (1,0,0) direction, (b) the (1,1,0) direction and (c) the (1,1,1) direction. Parameters are the same as in Fig.~\ref{fig:3d_envelopes}.}
\label{fig:3d_scatterplots}
\end{figure}

\end{widetext}

\bibliographystyle{apsrev4-1}
\bibliography{arxiv}

\begin{thebibliography}{46}%
\makeatletter
\providecommand \@ifxundefined [1]{%
 \@ifx{#1\undefined}
}%
\providecommand \@ifnum [1]{%
 \ifnum #1\expandafter \@firstoftwo
 \else \expandafter \@secondoftwo
 \fi
}%
\providecommand \@ifx [1]{%
 \ifx #1\expandafter \@firstoftwo
 \else \expandafter \@secondoftwo
 \fi
}%
\providecommand \natexlab [1]{#1}%
\providecommand \enquote  [1]{``#1''}%
\providecommand \bibnamefont  [1]{#1}%
\providecommand \bibfnamefont [1]{#1}%
\providecommand \citenamefont [1]{#1}%
\providecommand \href@noop [0]{\@secondoftwo}%
\providecommand \href [0]{\begingroup \@sanitize@url \@href}%
\providecommand \@href[1]{\@@startlink{#1}\@@href}%
\providecommand \@@href[1]{\endgroup#1\@@endlink}%
\providecommand \@sanitize@url [0]{\catcode `\\12\catcode `\$12\catcode
  `\&12\catcode `\#12\catcode `\^12\catcode `\_12\catcode `\%12\relax}%
\providecommand \@@startlink[1]{}%
\providecommand \@@endlink[0]{}%
\providecommand \url  [0]{\begingroup\@sanitize@url \@url }%
\providecommand \@url [1]{\endgroup\@href {#1}{\urlprefix }}%
\providecommand \urlprefix  [0]{URL }%
\providecommand \Eprint [0]{\href }%
\providecommand \doibase [0]{https://doi.org/}%
\providecommand \selectlanguage [0]{\@gobble}%
\providecommand \bibinfo  [0]{\@secondoftwo}%
\providecommand \bibfield  [0]{\@secondoftwo}%
\providecommand \translation [1]{[#1]}%
\providecommand \BibitemOpen [0]{}%
\providecommand \bibitemStop [0]{}%
\providecommand \bibitemNoStop [0]{.\EOS\space}%
\providecommand \EOS [0]{\spacefactor3000\relax}%
\providecommand \BibitemShut  [1]{\csname bibitem#1\endcsname}%
\let\auto@bib@innerbib\@empty
\bibitem [{\citenamefont {Greiner}\ \emph {et~al.}(2002)\citenamefont
  {Greiner}, \citenamefont {Mandel}, \citenamefont {Esslinger}, \citenamefont
  {H\"{a}nsch},\ and\ \citenamefont {Bloch}}]{Greiner2002}%
  \BibitemOpen
  \bibfield  {author} {\bibinfo {author} {\bibfnamefont {M.}~\bibnamefont
  {Greiner}}, \bibinfo {author} {\bibfnamefont {O.}~\bibnamefont {Mandel}},
  \bibinfo {author} {\bibfnamefont {T.}~\bibnamefont {Esslinger}}, \bibinfo
  {author} {\bibfnamefont {T.~W.}\ \bibnamefont {H\"{a}nsch}},\ and\ \bibinfo
  {author} {\bibfnamefont {I.}~\bibnamefont {Bloch}},\ }\href@noop {}
  {\bibfield  {journal} {\bibinfo  {journal} {Nature}\ }\textbf {\bibinfo
  {volume} {415}},\ \bibinfo {pages} {39} (\bibinfo {year} {2002})}\BibitemShut
  {NoStop}%
\bibitem [{\citenamefont {Bloch}(2005)}]{Bloch2005}%
  \BibitemOpen
  \bibfield  {author} {\bibinfo {author} {\bibfnamefont {I.}~\bibnamefont
  {Bloch}},\ }\href@noop {} {\bibfield  {journal} {\bibinfo  {journal} {Nat.
  Phys.}\ }\textbf {\bibinfo {volume} {1}},\ \bibinfo {pages} {23} (\bibinfo
  {year} {2005})}\BibitemShut {NoStop}%
\bibitem [{\citenamefont {Lewenstein}\ \emph {et~al.}(2007)\citenamefont
  {Lewenstein}, \citenamefont {Sanpera}, \citenamefont {Ahufinger},
  \citenamefont {Damski}, \citenamefont {Sen},\ and\ \citenamefont
  {Sen}}]{Lewenstein2007}%
  \BibitemOpen
  \bibfield  {author} {\bibinfo {author} {\bibfnamefont {M.}~\bibnamefont
  {Lewenstein}}, \bibinfo {author} {\bibfnamefont {A.}~\bibnamefont {Sanpera}},
  \bibinfo {author} {\bibfnamefont {V.}~\bibnamefont {Ahufinger}}, \bibinfo
  {author} {\bibfnamefont {B.}~\bibnamefont {Damski}}, \bibinfo {author}
  {\bibfnamefont {A.}~\bibnamefont {Sen}},\ and\ \bibinfo {author}
  {\bibfnamefont {U.}~\bibnamefont {Sen}},\ }\href@noop {} {\bibfield
  {journal} {\bibinfo  {journal} {Adv. Phys.}\ }\textbf {\bibinfo {volume}
  {56}},\ \bibinfo {pages} {243} (\bibinfo {year} {2007})}\BibitemShut
  {NoStop}%
\bibitem [{\citenamefont {Bloch}\ \emph {et~al.}(2008)\citenamefont {Bloch},
  \citenamefont {Dalibard},\ and\ \citenamefont {Zwerger}}]{Bloch2008}%
  \BibitemOpen
  \bibfield  {author} {\bibinfo {author} {\bibfnamefont {I.}~\bibnamefont
  {Bloch}}, \bibinfo {author} {\bibfnamefont {J.}~\bibnamefont {Dalibard}},\
  and\ \bibinfo {author} {\bibfnamefont {W.}~\bibnamefont {Zwerger}},\
  }\href@noop {} {\bibfield  {journal} {\bibinfo  {journal} {Rev. Mod. Phys.}\
  }\textbf {\bibinfo {volume} {80}},\ \bibinfo {pages} {885} (\bibinfo {year}
  {2008})}\BibitemShut {NoStop}%
\bibitem [{\citenamefont {Hung}\ \emph {et~al.}(2010)\citenamefont {Hung},
  \citenamefont {Zhang}, \citenamefont {Gemelke},\ and\ \citenamefont
  {Chin}}]{Hung2010}%
  \BibitemOpen
  \bibfield  {author} {\bibinfo {author} {\bibfnamefont {C.-L.}\ \bibnamefont
  {Hung}}, \bibinfo {author} {\bibfnamefont {X.}~\bibnamefont {Zhang}},
  \bibinfo {author} {\bibfnamefont {N.}~\bibnamefont {Gemelke}},\ and\ \bibinfo
  {author} {\bibfnamefont {C.}~\bibnamefont {Chin}},\ }\href@noop {} {\bibfield
   {journal} {\bibinfo  {journal} {Phys. Rev. Lett.}\ }\textbf {\bibinfo
  {volume} {104}},\ \bibinfo {pages} {160403} (\bibinfo {year}
  {2010})}\BibitemShut {NoStop}%
\bibitem [{\citenamefont {Chen}\ \emph {et~al.}(2011)\citenamefont {Chen},
  \citenamefont {White}, \citenamefont {Borries},\ and\ \citenamefont
  {DeMarco}}]{Chen2011}%
  \BibitemOpen
  \bibfield  {author} {\bibinfo {author} {\bibfnamefont {D.}~\bibnamefont
  {Chen}}, \bibinfo {author} {\bibfnamefont {M.}~\bibnamefont {White}},
  \bibinfo {author} {\bibfnamefont {C.}~\bibnamefont {Borries}},\ and\ \bibinfo
  {author} {\bibfnamefont {B.}~\bibnamefont {DeMarco}},\ }\href@noop {}
  {\bibfield  {journal} {\bibinfo  {journal} {Phys. Rev. Lett.}\ }\textbf
  {\bibinfo {volume} {106}},\ \bibinfo {pages} {235304} (\bibinfo {year}
  {2011})}\BibitemShut {NoStop}%
\bibitem [{\citenamefont {Kennett}(2013)}]{Kennett2013}%
  \BibitemOpen
  \bibfield  {author} {\bibinfo {author} {\bibfnamefont {M.~P.}\ \bibnamefont
  {Kennett}},\ }\href@noop {} {\bibfield  {journal} {\bibinfo  {journal} {ISRN
  Cond. Matter Phys.}\ }\textbf {\bibinfo {volume} {2013}},\ \bibinfo {pages}
  {393616} (\bibinfo {year} {2013})}\BibitemShut {NoStop}%
\bibitem [{\citenamefont {Gross}\ and\ \citenamefont
  {Bloch}(2017)}]{Gross2017}%
  \BibitemOpen
  \bibfield  {author} {\bibinfo {author} {\bibfnamefont {C.}~\bibnamefont
  {Gross}}\ and\ \bibinfo {author} {\bibfnamefont {I.}~\bibnamefont {Bloch}},\
  }\href@noop {} {\bibfield  {journal} {\bibinfo  {journal} {Science}\ }\textbf
  {\bibinfo {volume} {357}},\ \bibinfo {pages} {995} (\bibinfo {year}
  {2017})}\BibitemShut {NoStop}%
\bibitem [{\citenamefont {Fisher}\ \emph {et~al.}(1989)\citenamefont {Fisher},
  \citenamefont {Weichman}, \citenamefont {Grinstein},\ and\ \citenamefont
  {Fisher}}]{Fisher1989}%
  \BibitemOpen
  \bibfield  {author} {\bibinfo {author} {\bibfnamefont {M.~P.~A.}\
  \bibnamefont {Fisher}}, \bibinfo {author} {\bibfnamefont {P.~B.}\
  \bibnamefont {Weichman}}, \bibinfo {author} {\bibfnamefont {G.}~\bibnamefont
  {Grinstein}},\ and\ \bibinfo {author} {\bibfnamefont {D.~S.}\ \bibnamefont
  {Fisher}},\ }\href@noop {} {\bibfield  {journal} {\bibinfo  {journal} {Phys.
  Rev. B}\ }\textbf {\bibinfo {volume} {40}},\ \bibinfo {pages} {546} (\bibinfo
  {year} {1989})}\BibitemShut {NoStop}%
\bibitem [{\citenamefont {Jaksch}\ \emph {et~al.}(1998)\citenamefont {Jaksch},
  \citenamefont {Bruder}, \citenamefont {Cirac}, \citenamefont {Gardiner},\
  and\ \citenamefont {Zoller}}]{Jaksch1998}%
  \BibitemOpen
  \bibfield  {author} {\bibinfo {author} {\bibfnamefont {D.}~\bibnamefont
  {Jaksch}}, \bibinfo {author} {\bibfnamefont {C.}~\bibnamefont {Bruder}},
  \bibinfo {author} {\bibfnamefont {J.~I.}\ \bibnamefont {Cirac}}, \bibinfo
  {author} {\bibfnamefont {C.~W.}\ \bibnamefont {Gardiner}},\ and\ \bibinfo
  {author} {\bibfnamefont {P.}~\bibnamefont {Zoller}},\ }\href@noop {}
  {\bibfield  {journal} {\bibinfo  {journal} {Phys. Rev. Lett.}\ }\textbf
  {\bibinfo {volume} {81}},\ \bibinfo {pages} {3108} (\bibinfo {year}
  {1998})}\BibitemShut {NoStop}%
\bibitem [{\citenamefont {Choi}\ \emph {et~al.}(2016)\citenamefont {Choi},
  \citenamefont {Hild}, \citenamefont {Zeiher}, \citenamefont {Schau\ss},
  \citenamefont {Rubio-Abadal}, \citenamefont {Yefsah}, \citenamefont
  {Khemani}, \citenamefont {Huse}, \citenamefont {Bloch},\ and\ \citenamefont
  {Gross}}]{Choi2016}%
  \BibitemOpen
  \bibfield  {author} {\bibinfo {author} {\bibfnamefont {J.-Y.}\ \bibnamefont
  {Choi}}, \bibinfo {author} {\bibfnamefont {S.}~\bibnamefont {Hild}}, \bibinfo
  {author} {\bibfnamefont {J.}~\bibnamefont {Zeiher}}, \bibinfo {author}
  {\bibfnamefont {P.}~\bibnamefont {Schau\ss}}, \bibinfo {author}
  {\bibfnamefont {A.}~\bibnamefont {Rubio-Abadal}}, \bibinfo {author}
  {\bibfnamefont {T.}~\bibnamefont {Yefsah}}, \bibinfo {author} {\bibfnamefont
  {V.}~\bibnamefont {Khemani}}, \bibinfo {author} {\bibfnamefont {D.~A.}\
  \bibnamefont {Huse}}, \bibinfo {author} {\bibfnamefont {I.}~\bibnamefont
  {Bloch}},\ and\ \bibinfo {author} {\bibfnamefont {C.}~\bibnamefont {Gross}},\
  }\href@noop {} {\bibfield  {journal} {\bibinfo  {journal} {Science}\ }\textbf
  {\bibinfo {volume} {352}},\ \bibinfo {pages} {1547} (\bibinfo {year}
  {2016})}\BibitemShut {NoStop}%
\bibitem [{\citenamefont {Takasu}\ \emph {et~al.}(2020)\citenamefont {Takasu},
  \citenamefont {Yagami}, \citenamefont {Asaka}, \citenamefont {Fukushima},
  \citenamefont {Nagao}, \citenamefont {Goto}, \citenamefont {Danshita},\ and\
  \citenamefont {Takahashi}}]{Takasu2020}%
  \BibitemOpen
  \bibfield  {author} {\bibinfo {author} {\bibfnamefont {Y.}~\bibnamefont
  {Takasu}}, \bibinfo {author} {\bibfnamefont {T.}~\bibnamefont {Yagami}},
  \bibinfo {author} {\bibfnamefont {H.}~\bibnamefont {Asaka}}, \bibinfo
  {author} {\bibfnamefont {Y.}~\bibnamefont {Fukushima}}, \bibinfo {author}
  {\bibfnamefont {K.}~\bibnamefont {Nagao}}, \bibinfo {author} {\bibfnamefont
  {S.}~\bibnamefont {Goto}}, \bibinfo {author} {\bibfnamefont {I.}~\bibnamefont
  {Danshita}},\ and\ \bibinfo {author} {\bibfnamefont {Y.}~\bibnamefont
  {Takahashi}},\ }\href@noop {} {\bibfield  {journal} {\bibinfo  {journal}
  {Science Advances}\ }\textbf {\bibinfo {volume} {6}},\ \bibinfo {pages}
  {eaba9255} (\bibinfo {year} {2020})}\BibitemShut {NoStop}%
\bibitem [{\citenamefont {Bose}(2007)}]{Bose2007}%
  \BibitemOpen
  \bibfield  {author} {\bibinfo {author} {\bibfnamefont {S.}~\bibnamefont
  {Bose}},\ }\href@noop {} {\bibfield  {journal} {\bibinfo  {journal} {Contemp.
  Phys.}\ }\textbf {\bibinfo {volume} {48}},\ \bibinfo {pages} {13} (\bibinfo
  {year} {2007})}\BibitemShut {NoStop}%
\bibitem [{\citenamefont {Lieb}\ and\ \citenamefont
  {Robinson}(1972)}]{Lieb1972}%
  \BibitemOpen
  \bibfield  {author} {\bibinfo {author} {\bibfnamefont {E.~H.}\ \bibnamefont
  {Lieb}}\ and\ \bibinfo {author} {\bibfnamefont {D.~W.}\ \bibnamefont
  {Robinson}},\ }\href@noop {} {\bibfield  {journal} {\bibinfo  {journal}
  {Commun. Math. Phys.}\ }\textbf {\bibinfo {volume} {28}},\ \bibinfo {pages}
  {251} (\bibinfo {year} {1972})}\BibitemShut {NoStop}%
\bibitem [{\citenamefont {Nachtergaele}\ \emph {et~al.}(2009)\citenamefont
  {Nachtergaele}, \citenamefont {Raz}, \citenamefont {Schlein},\ and\
  \citenamefont {Sims}}]{Nachtergaele2009}%
  \BibitemOpen
  \bibfield  {author} {\bibinfo {author} {\bibfnamefont {B.}~\bibnamefont
  {Nachtergaele}}, \bibinfo {author} {\bibfnamefont {H.}~\bibnamefont {Raz}},
  \bibinfo {author} {\bibfnamefont {B.}~\bibnamefont {Schlein}},\ and\ \bibinfo
  {author} {\bibfnamefont {R.}~\bibnamefont {Sims}},\ }\href@noop {} {\bibfield
   {journal} {\bibinfo  {journal} {Commun. Math. Phys.}\ }\textbf {\bibinfo
  {volume} {286}},\ \bibinfo {pages} {1073} (\bibinfo {year}
  {2009})}\BibitemShut {NoStop}%
\bibitem [{\citenamefont {Eisert}\ and\ \citenamefont
  {Gross}(2009)}]{Eisert2009}%
  \BibitemOpen
  \bibfield  {author} {\bibinfo {author} {\bibfnamefont {J.}~\bibnamefont
  {Eisert}}\ and\ \bibinfo {author} {\bibfnamefont {D.}~\bibnamefont {Gross}},\
  }\href@noop {} {\bibfield  {journal} {\bibinfo  {journal} {Phys. Rev. Lett.}\
  }\textbf {\bibinfo {volume} {102}},\ \bibinfo {pages} {240501} (\bibinfo
  {year} {2009})}\BibitemShut {NoStop}%
\bibitem [{\citenamefont {J\"{u}nemann}\ \emph {et~al.}(2013)\citenamefont
  {J\"{u}nemann}, \citenamefont {Cadarso}, \citenamefont
  {P\'{e}rez-Garc\'{i}a}, \citenamefont {Bermudez},\ and\ \citenamefont
  {Garc\'{i}a-Ripoll}}]{Juneman2013}%
  \BibitemOpen
  \bibfield  {author} {\bibinfo {author} {\bibfnamefont {J.}~\bibnamefont
  {J\"{u}nemann}}, \bibinfo {author} {\bibfnamefont {A.}~\bibnamefont
  {Cadarso}}, \bibinfo {author} {\bibfnamefont {D.}~\bibnamefont
  {P\'{e}rez-Garc\'{i}a}}, \bibinfo {author} {\bibfnamefont {A.}~\bibnamefont
  {Bermudez}},\ and\ \bibinfo {author} {\bibfnamefont {J.~J.}\ \bibnamefont
  {Garc\'{i}a-Ripoll}},\ }\href@noop {} {\bibfield  {journal} {\bibinfo
  {journal} {Phys. Rev. Lett.}\ }\textbf {\bibinfo {volume} {111}},\ \bibinfo
  {pages} {230404} (\bibinfo {year} {2013})}\BibitemShut {NoStop}%
\bibitem [{\citenamefont {Schuch}\ \emph {et~al.}(2011)\citenamefont {Schuch},
  \citenamefont {Harrison}, \citenamefont {Osborne},\ and\ \citenamefont
  {Eisert}}]{Schuch2011}%
  \BibitemOpen
  \bibfield  {author} {\bibinfo {author} {\bibfnamefont {N.}~\bibnamefont
  {Schuch}}, \bibinfo {author} {\bibfnamefont {S.~K.}\ \bibnamefont
  {Harrison}}, \bibinfo {author} {\bibfnamefont {T.~J.}\ \bibnamefont
  {Osborne}},\ and\ \bibinfo {author} {\bibfnamefont {J.}~\bibnamefont
  {Eisert}},\ }\href@noop {} {\bibfield  {journal} {\bibinfo  {journal} {Phys.
  Rev. A}\ }\textbf {\bibinfo {volume} {84}},\ \bibinfo {pages} {032309}
  (\bibinfo {year} {2011})}\BibitemShut {NoStop}%
\bibitem [{\citenamefont {Bakr}\ \emph {et~al.}(2010)\citenamefont {Bakr},
  \citenamefont {Peng}, \citenamefont {Tai}, \citenamefont {Ma}, \citenamefont
  {Simon}, \citenamefont {Gillen}, \citenamefont {F\"{o}lling}, \citenamefont
  {Pollet},\ and\ \citenamefont {Greiner}}]{Bakr2010}%
  \BibitemOpen
  \bibfield  {author} {\bibinfo {author} {\bibfnamefont {W.~S.}\ \bibnamefont
  {Bakr}}, \bibinfo {author} {\bibfnamefont {A.}~\bibnamefont {Peng}}, \bibinfo
  {author} {\bibfnamefont {M.~E.}\ \bibnamefont {Tai}}, \bibinfo {author}
  {\bibfnamefont {R.}~\bibnamefont {Ma}}, \bibinfo {author} {\bibfnamefont
  {J.}~\bibnamefont {Simon}}, \bibinfo {author} {\bibfnamefont {J.~I.}\
  \bibnamefont {Gillen}}, \bibinfo {author} {\bibfnamefont {S.}~\bibnamefont
  {F\"{o}lling}}, \bibinfo {author} {\bibfnamefont {L.}~\bibnamefont
  {Pollet}},\ and\ \bibinfo {author} {\bibfnamefont {M.}~\bibnamefont
  {Greiner}},\ }\href@noop {} {\bibfield  {journal} {\bibinfo  {journal}
  {Science}\ }\textbf {\bibinfo {volume} {329}},\ \bibinfo {pages} {547}
  (\bibinfo {year} {2010})}\BibitemShut {NoStop}%
\bibitem [{\citenamefont {Sherson}\ \emph {et~al.}(2010)\citenamefont
  {Sherson}, \citenamefont {Weitenberg}, \citenamefont {Endres}, \citenamefont
  {Cheneau}, \citenamefont {Bloch},\ and\ \citenamefont {Kuhr}}]{Sherson2010}%
  \BibitemOpen
  \bibfield  {author} {\bibinfo {author} {\bibfnamefont {J.~F.}\ \bibnamefont
  {Sherson}}, \bibinfo {author} {\bibfnamefont {C.}~\bibnamefont {Weitenberg}},
  \bibinfo {author} {\bibfnamefont {M.}~\bibnamefont {Endres}}, \bibinfo
  {author} {\bibfnamefont {M.}~\bibnamefont {Cheneau}}, \bibinfo {author}
  {\bibfnamefont {I.}~\bibnamefont {Bloch}},\ and\ \bibinfo {author}
  {\bibfnamefont {S.}~\bibnamefont {Kuhr}},\ }\href@noop {} {\bibfield
  {journal} {\bibinfo  {journal} {Nature}\ }\textbf {\bibinfo {volume} {467}},\
  \bibinfo {pages} {68} (\bibinfo {year} {2010})}\BibitemShut {NoStop}%
\bibitem [{\citenamefont {Cheneau}\ \emph {et~al.}(2012)\citenamefont
  {Cheneau}, \citenamefont {Barmettler}, \citenamefont {Poletti}, \citenamefont
  {Endres}, \citenamefont {Schau\ss}, \citenamefont {Fukuhara}, \citenamefont
  {Gross}, \citenamefont {Bloch}, \citenamefont {Kollath},\ and\ \citenamefont
  {Kuhr}}]{Cheneau2012}%
  \BibitemOpen
  \bibfield  {author} {\bibinfo {author} {\bibfnamefont {M.}~\bibnamefont
  {Cheneau}}, \bibinfo {author} {\bibfnamefont {P.}~\bibnamefont {Barmettler}},
  \bibinfo {author} {\bibfnamefont {D.}~\bibnamefont {Poletti}}, \bibinfo
  {author} {\bibfnamefont {M.}~\bibnamefont {Endres}}, \bibinfo {author}
  {\bibfnamefont {P.}~\bibnamefont {Schau\ss}}, \bibinfo {author}
  {\bibfnamefont {T.}~\bibnamefont {Fukuhara}}, \bibinfo {author}
  {\bibfnamefont {C.}~\bibnamefont {Gross}}, \bibinfo {author} {\bibfnamefont
  {I.}~\bibnamefont {Bloch}}, \bibinfo {author} {\bibfnamefont
  {C.}~\bibnamefont {Kollath}},\ and\ \bibinfo {author} {\bibfnamefont
  {S.}~\bibnamefont {Kuhr}},\ }\href@noop {} {\bibfield  {journal} {\bibinfo
  {journal} {Nature}\ }\textbf {\bibinfo {volume} {481}},\ \bibinfo {pages}
  {484} (\bibinfo {year} {2012})}\BibitemShut {NoStop}%
\bibitem [{\citenamefont {Clark}\ and\ \citenamefont
  {Jaksch}(2004)}]{Clark2004}%
  \BibitemOpen
  \bibfield  {author} {\bibinfo {author} {\bibfnamefont {S.~R.}\ \bibnamefont
  {Clark}}\ and\ \bibinfo {author} {\bibfnamefont {D.}~\bibnamefont {Jaksch}},\
  }\href@noop {} {\bibfield  {journal} {\bibinfo  {journal} {Phys. Rev. A}\
  }\textbf {\bibinfo {volume} {70}},\ \bibinfo {pages} {043612} (\bibinfo
  {year} {2004})}\BibitemShut {NoStop}%
\bibitem [{\citenamefont {Kollath}\ \emph {et~al.}(2007)\citenamefont
  {Kollath}, \citenamefont {L\"{a}uchli},\ and\ \citenamefont
  {Altman}}]{Kollath2007}%
  \BibitemOpen
  \bibfield  {author} {\bibinfo {author} {\bibfnamefont {C.}~\bibnamefont
  {Kollath}}, \bibinfo {author} {\bibfnamefont {A.~M.}\ \bibnamefont
  {L\"{a}uchli}},\ and\ \bibinfo {author} {\bibfnamefont {E.}~\bibnamefont
  {Altman}},\ }\href@noop {} {\bibfield  {journal} {\bibinfo  {journal} {Phys.
  Rev. Lett.}\ }\textbf {\bibinfo {volume} {98}},\ \bibinfo {pages} {180601}
  (\bibinfo {year} {2007})}\BibitemShut {NoStop}%
\bibitem [{\citenamefont {L\"{a}uchli}\ and\ \citenamefont
  {Kollath}(2008)}]{Lauchli2008}%
  \BibitemOpen
  \bibfield  {author} {\bibinfo {author} {\bibfnamefont {A.~M.}\ \bibnamefont
  {L\"{a}uchli}}\ and\ \bibinfo {author} {\bibfnamefont {C.}~\bibnamefont
  {Kollath}},\ }\href@noop {} {\bibfield  {journal} {\bibinfo  {journal} {J.
  Stat. Mech.}\ }\textbf {\bibinfo {volume} {2008}},\ \bibinfo {pages} {P05018}
  (\bibinfo {year} {2008})}\BibitemShut {NoStop}%
\bibitem [{\citenamefont {Bernier}\ \emph {et~al.}(2011)\citenamefont
  {Bernier}, \citenamefont {Roux},\ and\ \citenamefont
  {Kollath}}]{Bernier2011}%
  \BibitemOpen
  \bibfield  {author} {\bibinfo {author} {\bibfnamefont {J.-S.}\ \bibnamefont
  {Bernier}}, \bibinfo {author} {\bibfnamefont {G.}~\bibnamefont {Roux}},\ and\
  \bibinfo {author} {\bibfnamefont {C.}~\bibnamefont {Kollath}},\ }\href@noop
  {} {\bibfield  {journal} {\bibinfo  {journal} {Phys. Rev. Lett.}\ }\textbf
  {\bibinfo {volume} {106}},\ \bibinfo {pages} {200601} (\bibinfo {year}
  {2011})}\BibitemShut {NoStop}%
\bibitem [{\citenamefont {Bernier}\ \emph {et~al.}(2012)\citenamefont
  {Bernier}, \citenamefont {Poletti}, \citenamefont {Barmettler}, \citenamefont
  {Roux},\ and\ \citenamefont {Kollath}}]{Bernier2012}%
  \BibitemOpen
  \bibfield  {author} {\bibinfo {author} {\bibfnamefont {J.-S.}\ \bibnamefont
  {Bernier}}, \bibinfo {author} {\bibfnamefont {D.}~\bibnamefont {Poletti}},
  \bibinfo {author} {\bibfnamefont {P.}~\bibnamefont {Barmettler}}, \bibinfo
  {author} {\bibfnamefont {G.}~\bibnamefont {Roux}},\ and\ \bibinfo {author}
  {\bibfnamefont {C.}~\bibnamefont {Kollath}},\ }\href@noop {} {\bibfield
  {journal} {\bibinfo  {journal} {Phys. Rev. A}\ }\textbf {\bibinfo {volume}
  {85}},\ \bibinfo {pages} {033641} (\bibinfo {year} {2012})}\BibitemShut
  {NoStop}%
\bibitem [{\citenamefont {Barmettler}\ \emph {et~al.}(2012)\citenamefont
  {Barmettler}, \citenamefont {Poletti}, \citenamefont {Cheneau},\ and\
  \citenamefont {Kollath}}]{Barmettler2012}%
  \BibitemOpen
  \bibfield  {author} {\bibinfo {author} {\bibfnamefont {P.}~\bibnamefont
  {Barmettler}}, \bibinfo {author} {\bibfnamefont {D.}~\bibnamefont {Poletti}},
  \bibinfo {author} {\bibfnamefont {M.}~\bibnamefont {Cheneau}},\ and\ \bibinfo
  {author} {\bibfnamefont {C.}~\bibnamefont {Kollath}},\ }\href@noop {}
  {\bibfield  {journal} {\bibinfo  {journal} {Phys. Rev. A}\ }\textbf {\bibinfo
  {volume} {85}},\ \bibinfo {pages} {053625} (\bibinfo {year}
  {2012})}\BibitemShut {NoStop}%
\bibitem [{\citenamefont {Trotzky}\ \emph {et~al.}(2012)\citenamefont
  {Trotzky}, \citenamefont {Chen}, \citenamefont {Flesch}, \citenamefont
  {McCulloch}, \citenamefont {Schollw\"{o}ck}, \citenamefont {Eisert},\ and\
  \citenamefont {Bloch}}]{Trotzky2012}%
  \BibitemOpen
  \bibfield  {author} {\bibinfo {author} {\bibfnamefont {S.}~\bibnamefont
  {Trotzky}}, \bibinfo {author} {\bibfnamefont {Y.-A.}\ \bibnamefont {Chen}},
  \bibinfo {author} {\bibfnamefont {A.}~\bibnamefont {Flesch}}, \bibinfo
  {author} {\bibfnamefont {I.~P.}\ \bibnamefont {McCulloch}}, \bibinfo {author}
  {\bibfnamefont {U.}~\bibnamefont {Schollw\"{o}ck}}, \bibinfo {author}
  {\bibfnamefont {J.}~\bibnamefont {Eisert}},\ and\ \bibinfo {author}
  {\bibfnamefont {I.}~\bibnamefont {Bloch}},\ }\href@noop {} {\bibfield
  {journal} {\bibinfo  {journal} {Nat. Phys.}\ }\textbf {\bibinfo {volume}
  {8}},\ \bibinfo {pages} {325} (\bibinfo {year} {2012})}\BibitemShut {NoStop}%
\bibitem [{\citenamefont {Cevolani}\ \emph {et~al.}(2018)\citenamefont
  {Cevolani}, \citenamefont {Despres}, \citenamefont {Carleo}, \citenamefont
  {Tagliacozzo},\ and\ \citenamefont {Sanchez-Palencia}}]{Cevolani2018}%
  \BibitemOpen
  \bibfield  {author} {\bibinfo {author} {\bibfnamefont {L.}~\bibnamefont
  {Cevolani}}, \bibinfo {author} {\bibfnamefont {J.}~\bibnamefont {Despres}},
  \bibinfo {author} {\bibfnamefont {G.}~\bibnamefont {Carleo}}, \bibinfo
  {author} {\bibfnamefont {L.}~\bibnamefont {Tagliacozzo}},\ and\ \bibinfo
  {author} {\bibfnamefont {L.}~\bibnamefont {Sanchez-Palencia}},\ }\href@noop
  {} {\bibfield  {journal} {\bibinfo  {journal} {Phys. Rev. B}\ }\textbf
  {\bibinfo {volume} {98}},\ \bibinfo {pages} {024302} (\bibinfo {year}
  {2018})}\BibitemShut {NoStop}%
\bibitem [{\citenamefont {Despres}\ \emph {et~al.}(2019)\citenamefont
  {Despres}, \citenamefont {Villa},\ and\ \citenamefont
  {Sanchez-Palencia}}]{Despres2019}%
  \BibitemOpen
  \bibfield  {author} {\bibinfo {author} {\bibfnamefont {J.}~\bibnamefont
  {Despres}}, \bibinfo {author} {\bibfnamefont {L.}~\bibnamefont {Villa}},\
  and\ \bibinfo {author} {\bibfnamefont {L.}~\bibnamefont {Sanchez-Palencia}},\
  }\href@noop {} {\bibfield  {journal} {\bibinfo  {journal} {Sci. Rep.}\
  }\textbf {\bibinfo {volume} {9}},\ \bibinfo {pages} {4135} (\bibinfo {year}
  {2019})}\BibitemShut {NoStop}%
\bibitem [{\citenamefont {Navez}\ and\ \citenamefont
  {Sch\"{u}tzhold}(2010)}]{Navez2010}%
  \BibitemOpen
  \bibfield  {author} {\bibinfo {author} {\bibfnamefont {P.}~\bibnamefont
  {Navez}}\ and\ \bibinfo {author} {\bibfnamefont {R.}~\bibnamefont
  {Sch\"{u}tzhold}},\ }\href@noop {} {\bibfield  {journal} {\bibinfo  {journal}
  {Phys. Rev. A}\ }\textbf {\bibinfo {volume} {82}},\ \bibinfo {pages} {063603}
  (\bibinfo {year} {2010})}\BibitemShut {NoStop}%
\bibitem [{\citenamefont {Trefzger}\ and\ \citenamefont
  {Sengupta}(2011)}]{Trefzger2011}%
  \BibitemOpen
  \bibfield  {author} {\bibinfo {author} {\bibfnamefont {C.}~\bibnamefont
  {Trefzger}}\ and\ \bibinfo {author} {\bibfnamefont {K.}~\bibnamefont
  {Sengupta}},\ }\href@noop {} {\bibfield  {journal} {\bibinfo  {journal}
  {Phys. Rev. Lett.}\ }\textbf {\bibinfo {volume} {106}},\ \bibinfo {pages}
  {095702} (\bibinfo {year} {2011})}\BibitemShut {NoStop}%
\bibitem [{\citenamefont {Krutitsky}\ \emph {et~al.}(2014)\citenamefont
  {Krutitsky}, \citenamefont {Navez}, \citenamefont {Quiesser},\ and\
  \citenamefont {Sch\"{u}tzhold}}]{Krutitsky2014}%
  \BibitemOpen
  \bibfield  {author} {\bibinfo {author} {\bibfnamefont {K.~V.}\ \bibnamefont
  {Krutitsky}}, \bibinfo {author} {\bibfnamefont {P.}~\bibnamefont {Navez}},
  \bibinfo {author} {\bibfnamefont {F.}~\bibnamefont {Quiesser}},\ and\
  \bibinfo {author} {\bibfnamefont {R.}~\bibnamefont {Sch\"{u}tzhold}},\
  }\href@noop {} {\bibfield  {journal} {\bibinfo  {journal} {Eur. Phys. J.
  Quant. Tech.}\ }\textbf {\bibinfo {volume} {1}},\ \bibinfo {pages} {12}
  (\bibinfo {year} {2014})}\BibitemShut {NoStop}%
\bibitem [{\citenamefont {Queisser}\ \emph {et~al.}(2014)\citenamefont
  {Queisser}, \citenamefont {Krutitsky}, \citenamefont {Navez},\ and\
  \citenamefont {Sch\"{u}tzhold}}]{Quiesser2014}%
  \BibitemOpen
  \bibfield  {author} {\bibinfo {author} {\bibfnamefont {F.}~\bibnamefont
  {Queisser}}, \bibinfo {author} {\bibfnamefont {K.~V.}\ \bibnamefont
  {Krutitsky}}, \bibinfo {author} {\bibfnamefont {P.}~\bibnamefont {Navez}},\
  and\ \bibinfo {author} {\bibfnamefont {R.}~\bibnamefont {Sch\"{u}tzhold}},\
  }\href@noop {} {\bibfield  {journal} {\bibinfo  {journal} {Phys. Rev. A}\
  }\textbf {\bibinfo {volume} {89}},\ \bibinfo {pages} {033616} (\bibinfo
  {year} {2014})}\BibitemShut {NoStop}%
\bibitem [{\citenamefont {Carleo}\ \emph {et~al.}(2014)\citenamefont {Carleo},
  \citenamefont {Becca}, \citenamefont {Sanchez-Palencia}, \citenamefont
  {Sorella},\ and\ \citenamefont {Fabrizio}}]{Carleo2014}%
  \BibitemOpen
  \bibfield  {author} {\bibinfo {author} {\bibfnamefont {G.}~\bibnamefont
  {Carleo}}, \bibinfo {author} {\bibfnamefont {F.}~\bibnamefont {Becca}},
  \bibinfo {author} {\bibfnamefont {L.}~\bibnamefont {Sanchez-Palencia}},
  \bibinfo {author} {\bibfnamefont {S.}~\bibnamefont {Sorella}},\ and\ \bibinfo
  {author} {\bibfnamefont {M.}~\bibnamefont {Fabrizio}},\ }\href@noop {}
  {\bibfield  {journal} {\bibinfo  {journal} {Phys. Rev. A}\ }\textbf {\bibinfo
  {volume} {89}},\ \bibinfo {pages} {031602(R)} (\bibinfo {year}
  {2014})}\BibitemShut {NoStop}%
\bibitem [{\citenamefont {Yanay}\ and\ \citenamefont
  {Mueller}(2016)}]{Yanay2016}%
  \BibitemOpen
  \bibfield  {author} {\bibinfo {author} {\bibfnamefont {Y.}~\bibnamefont
  {Yanay}}\ and\ \bibinfo {author} {\bibfnamefont {E.~J.}\ \bibnamefont
  {Mueller}},\ }\href@noop {} {\bibfield  {journal} {\bibinfo  {journal} {Phys.
  Rev. A}\ }\textbf {\bibinfo {volume} {93}},\ \bibinfo {pages} {013622}
  (\bibinfo {year} {2016})}\BibitemShut {NoStop}%
\bibitem [{\citenamefont {Cornwall}\ \emph {et~al.}(1974)\citenamefont
  {Cornwall}, \citenamefont {Jackiw},\ and\ \citenamefont
  {Tomboulis}}]{Cornwall1974}%
  \BibitemOpen
  \bibfield  {author} {\bibinfo {author} {\bibfnamefont {J.~M.}\ \bibnamefont
  {Cornwall}}, \bibinfo {author} {\bibfnamefont {R.}~\bibnamefont {Jackiw}},\
  and\ \bibinfo {author} {\bibfnamefont {E.}~\bibnamefont {Tomboulis}},\
  }\href@noop {} {\bibfield  {journal} {\bibinfo  {journal} {Phys. Rev. D}\
  }\textbf {\bibinfo {volume} {10}},\ \bibinfo {pages} {2428} (\bibinfo {year}
  {1974})}\BibitemShut {NoStop}%
\bibitem [{\citenamefont {Berges}(2004)}]{Berges2004}%
  \BibitemOpen
  \bibfield  {author} {\bibinfo {author} {\bibfnamefont {J.}~\bibnamefont
  {Berges}},\ }\href@noop {} {\bibfield  {journal} {\bibinfo  {journal} {AIP
  Conf. Proc.}\ }\textbf {\bibinfo {volume} {739}},\ \bibinfo {pages} {3}
  (\bibinfo {year} {2004})}\BibitemShut {NoStop}%
\bibitem [{\citenamefont {Kennett}\ and\ \citenamefont
  {Dalidovich}(2011)}]{Kennett2011}%
  \BibitemOpen
  \bibfield  {author} {\bibinfo {author} {\bibfnamefont {M.~P.}\ \bibnamefont
  {Kennett}}\ and\ \bibinfo {author} {\bibfnamefont {D.}~\bibnamefont
  {Dalidovich}},\ }\href@noop {} {\bibfield  {journal} {\bibinfo  {journal}
  {Phys. Rev. A}\ }\textbf {\bibinfo {volume} {84}},\ \bibinfo {pages} {033620}
  (\bibinfo {year} {2011})}\BibitemShut {NoStop}%
\bibitem [{\citenamefont {Fitzpatrick}\ and\ \citenamefont
  {Kennett}(2018{\natexlab{a}})}]{Fitzpatrick2018a}%
  \BibitemOpen
  \bibfield  {author} {\bibinfo {author} {\bibfnamefont {M.~R.~C.}\
  \bibnamefont {Fitzpatrick}}\ and\ \bibinfo {author} {\bibfnamefont {M.~P.}\
  \bibnamefont {Kennett}},\ }\href@noop {} {\bibfield  {journal} {\bibinfo
  {journal} {Nucl. Phys. B}\ }\textbf {\bibinfo {volume} {930}},\ \bibinfo
  {pages} {1} (\bibinfo {year} {2018}{\natexlab{a}})}\BibitemShut {NoStop}%
\bibitem [{\citenamefont {Fitzpatrick}\ and\ \citenamefont
  {Kennett}(2018{\natexlab{b}})}]{Fitzpatrick2018b}%
  \BibitemOpen
  \bibfield  {author} {\bibinfo {author} {\bibfnamefont {M.~R.~C.}\
  \bibnamefont {Fitzpatrick}}\ and\ \bibinfo {author} {\bibfnamefont {M.~P.}\
  \bibnamefont {Kennett}},\ }\href@noop {} {\bibfield  {journal} {\bibinfo
  {journal} {Phys. Rev. A}\ }\textbf {\bibinfo {volume} {98}},\ \bibinfo
  {pages} {053618} (\bibinfo {year} {2018}{\natexlab{b}})}\BibitemShut
  {NoStop}%
\bibitem [{\citenamefont {Kennett}\ and\ \citenamefont
  {Fitzpatrick}(2020)}]{Kennett2020}%
  \BibitemOpen
  \bibfield  {author} {\bibinfo {author} {\bibfnamefont {M.~P.}\ \bibnamefont
  {Kennett}}\ and\ \bibinfo {author} {\bibfnamefont {M.~R.~C.}\ \bibnamefont
  {Fitzpatrick}},\ }\href@noop {} {\bibfield  {journal} {\bibinfo  {journal}
  {J. Low Temp. Phys.}\ }\textbf {\bibinfo {volume} {201}},\ \bibinfo {pages}
  {82} (\bibinfo {year} {2020})}\BibitemShut {NoStop}%
\bibitem [{\citenamefont {Stefanucci}\ and\ \citenamefont {van
  Leeuwen}(2013)}]{Stefanucci2013}%
  \BibitemOpen
  \bibfield  {author} {\bibinfo {author} {\bibfnamefont {G.}~\bibnamefont
  {Stefanucci}}\ and\ \bibinfo {author} {\bibfnamefont {R.}~\bibnamefont {van
  Leeuwen}},\ }\href@noop {} {\emph {\bibinfo {title} {Nonequilibrium Many-Body
  Theory of Quantum Systems}}}\ (\bibinfo  {publisher} {Cambridge University
  Press},\ \bibinfo {address} {New York, NY},\ \bibinfo {year}
  {2013})\BibitemShut {NoStop}%
\bibitem [{\citenamefont {Fitzpatrick}(2019)}]{Fitzpatrick2019}%
  \BibitemOpen
  \bibfield  {author} {\bibinfo {author} {\bibfnamefont {M.~R.~C.}\
  \bibnamefont {Fitzpatrick}},\ }\href@noop {} {Ph.D. thesis},\ \bibinfo
  {school} {Simon Fraser University} (\bibinfo {year} {2019})\BibitemShut
  {NoStop}%
\bibitem [{\citenamefont {Konstantinov}\ and\ \citenamefont
  {Perel}(1961)}]{Konstantinov1961}%
  \BibitemOpen
  \bibfield  {author} {\bibinfo {author} {\bibfnamefont {O.~V.}\ \bibnamefont
  {Konstantinov}}\ and\ \bibinfo {author} {\bibfnamefont {V.~I.}\ \bibnamefont
  {Perel}},\ }\href@noop {} {\bibfield  {journal} {\bibinfo  {journal} {Zh.
  Eksp. Teor. Fiz.}\ }\textbf {\bibinfo {volume} {39}},\ \bibinfo {pages} {197}
  (\bibinfo {year} {1961})},\ \bibinfo {note} {[Sov. Phys. JETP {\bf 12}, 142
  (1961)]}\BibitemShut {NoStop}%
\bibitem [{\citenamefont {Sengupta}\ and\ \citenamefont
  {Dupuis}(2005)}]{Sengupta2005}%
  \BibitemOpen
  \bibfield  {author} {\bibinfo {author} {\bibfnamefont {K.}~\bibnamefont
  {Sengupta}}\ and\ \bibinfo {author} {\bibfnamefont {N.}~\bibnamefont
  {Dupuis}},\ }\href@noop {} {\bibfield  {journal} {\bibinfo  {journal} {Phys.
  Rev. A}\ }\textbf {\bibinfo {volume} {71}},\ \bibinfo {pages} {033629}
  (\bibinfo {year} {2005})}\BibitemShut {NoStop}%
\end{thebibliography}%

\end{document}